\documentclass[11pt]{article}

\usepackage{multicol,caption}
\usepackage{lipsum}
\newenvironment{Figure}
  {\par\medskip\noindent\minipage{\linewidth}}
  {\endminipage\par\medskip}

\usepackage{times}
\usepackage{amsfonts,amssymb,bm}
\usepackage{amsmath}
\usepackage{graphicx}
\usepackage{epsfig}
\usepackage{mathrsfs}
\usepackage{color,soul}
\usepackage{bbold}
\usepackage[english]{babel}
\usepackage{cite}
\usepackage{lipsum} 
\usepackage{hyperref}
\hypersetup{colorlinks,%
citecolor=black,%
filecolor=black,%
linkcolor=black,%
urlcolor=black}
\usepackage{pstricks}
\usepackage{multirow}
\usepackage{graphicx}
\usepackage{mathtools}
\usepackage{textcomp}
\usepackage{braket}
\usepackage{empheq}
\usepackage{color}
\usepackage{xcolor}
\usepackage{longtable}
\usepackage{booktabs,siunitx}
\usepackage{rotating}
\usepackage{mparhack}
\usepackage{amsmath}
\usepackage{eurosym}
\usepackage[babel]{csquotes}

\usepackage[font=small]{caption}
\usepackage{url}
\usepackage{hyperref}

\usepackage{newfloat}
\DeclareFloatingEnvironment[name={Supplementary Figure}]{suppfigure}

\definecolor{maroon}{rgb}{0.7,0,0}

\definecolor{ngreen}{rgb}{0.2,0.6,0.2}

\newcommand{\gra}{\color{gray}}
\definecolor{golden}{rgb}{0.8,0.6,0.1}

\newcommand{\beq}{\begin{equation}}
\newcommand{\eeq}{\end{equation}}
\newcommand{\bqa}{\begin{eqnarray}}
\newcommand{\eqa}{\end{eqnarray}}

\newcommand{\degree}{$^{\circ}$}
\DeclareMathOperator{\tr}{Tr}

\graphicspath{{../figures/}}

\newcounter{lastnote}

\title{Experimental Verification of an Indefinite Causal Order}

\author
{Giulia Rubino$^{1\ast}$, Lee A. Rozema$^{1}$, Adrien Feix$^{1,2}$, Mateus Ara\'ujo$^{1,2}$, Jonas M. Zeuner$^{1}$,\\ Lorenzo M. Procopio$^{1}$, \v{C}aslav Brukner$^{1,2}$, Philip Walther$^{1\ast}$\\
\\
\normalsize{$^{1}$Quantum Optics, -Nanophysics, -Information Group}, 
\normalsize{Faculty of Physics, University of Vienna,}\\ 
\normalsize{Boltzmanngasse 5, Vienna A-1090, Austria}\\
\normalsize{$^{2}$Institute for Quantum Optics \& Quantum Information (IQOQI)},
\normalsize{Austrian Academy of Sciences,}\\ 
\normalsize{Boltzmanngasse 3, Vienna A-1090, Austria}\\
\\
\normalsize{$^\ast$To whom correspondence should be addressed;}\\
\normalsize{ E-mail:  giulia.rubino@univie.ac.at (G.R.); philip.walther@univie.ac.at (P.W.)}
}

\begin{document} 


\baselineskip12pt


\maketitle

\date{}


\begin{abstract}
Investigating the role of causal order in quantum mechanics has recently revealed that the causal distribution of events may not be a-priori well-defined in quantum theory.
While this has triggered a growing interest on the theoretical side, creating processes without a causal order is an experimental task.
Here we report the first decisive demonstration of a process with an indefinite causal order.
To do this, we quantify how incompatible our set-up is with a definite causal order by measuring a `causal witness'.
This mathematical object incorporates a series of measurements which are designed to yield a certain outcome only if the process under examination is not consistent with \textit{any} well-defined causal order.
In our experiment we perform a measurement in a superposition of causal orders---without destroying the coherence---to acquire information both inside and outside of a `causally non-ordered process'.
Using this information, we experimentally determine a causal witness, demonstrating by almost seven standard deviations that the experimentally implemented process does not have a definite causal order.
\end{abstract}

\begin{multicols}{2}

The notion of causality is an innate concept, which defines the link between physical phenomena that temporally follow one another, one manifestly being cause of the other. Nevertheless, in quantum mechanics (QM) the concept of causality is not as straightforward. For example, when the superposition principle is applied to causal relations, situations without a definite causal-order can arise \cite{PhysRevA.88.022318}. Although this can lead to disconcerting consequences, forcing one to question concepts that are commonly viewed as the main ingredients of our physical description of the world \cite{NatPhys.10.259}, these effects can be exploited to achieve improvements in computational complexity \cite{Hardy2009,PhysRevA.86.040301,PhysRevLett.113.250402} and quantum communications \cite{PhysRevA.92.052326,Philippe,6874888}. Recently, this computational advantage was experimentally demonstrated in \cite{NatCommun.6}. However, the absence of a causal order was inferred from the success of an algorithm rather than being directly measured. In the present work, we explicitly demonstrate the realization of a causally non-ordered process by measuring a so-called `\textit{causal witness}' \cite{NewJournPhys.17.102001}. 

In order to make our results stronger (\textit{i.e.}, make the causal witness more robust to noise), we performed a superposition of the orders of a unitary gate and a measurement operation. In other words, we made a measurement \textit{inside} a quantum process with an indefinite order of operations (the quantum SWITCH \cite{PhysRevA.88.022318}). Performing a standard measurement inside the quantum SWITCH would destroy its coherence, since it would reveal the time at which the  measurement is performed, and would thus also reveal whether it is performed before or after other operations. In other words, such a measurement would reveal the causal order between the operations. In our scheme, however, the measurement outcomes are read out only ``at the end'' of the process, preserving its coherence. Since applications of indefinite causal orders will most likely require the superposition of orders of complex quantum operations, we believe that, in addition to the first direct demonstration of an indefinite causal order, our measurement in a quantum SWITCH can also be considered a technological step towards such applications \cite{Hardy2009,PhysRevA.86.040301,PhysRevLett.113.250402,PhysRevA.92.052326,Philippe,6874888}.
 
In our usual understanding of causal relations, if we consider two events $A$ and $B$ which are connected by a time-like vector, we will have \textit{one} of two cases: either `$A$ is in the past of $B$', or `$B$ is in the past of $A$'. However, the application of the superposition principle to such causal relations calls into question the interpretation of causal orders as a pre-existing property. In fact, the causal order can become genuinely indefinite. To see this, consider a two-qubit quantum state $\ket{\phi}$ lying in the composite Hilbert space $\mathcal{H}^C \otimes \mathcal{H}^T$ with $\mathcal{H}^C$ and $\mathcal{H}^T$ each being two-dimensional Hilbert spaces. It is possible to condition the order in which operations are applied to a target state $\ket{\psi}^T \in \mathcal{H}^T$ on the value of a control state $\ket{\chi}^C \in \mathcal{H}_C$. For example, if the state of the control qubit is $\ket{0}^C$, the two operators will be applied in the order ${A}$ and then ${B}$ on the state of the target qubit $\ket{\psi}^T$, and vice versa if the state of the control qubit is $\ket{1}^C$. Therefore, if the control qubit is in a superposition state $ \frac{1}{\sqrt{2}}\bigl(\ket{0}+\ket{1}\bigr)^C$, a controlled quantum superposition of the situations `${A}$ is in the past of ${B}$' and `${B}$ is in the past of ${A}$' is established (Figure \ref{img:quantum_switch}).

From this simple example we can see that the causal order between events is not always definite in QM. One could, in the spirit of hidden-variable theories, insist that there might anyway be a well-defined causal order. However, this claim requires, in general, a theory to be non-local and contextual because of the Bell and Kochen-Specker theorems \cite{Phys.1.195,JMathMech.2.59} and there is no reason to suppose that such a no-go theorem would also apply to \textit{causal orders} \cite{mag}.

\begin{Figure}
\centering
\includegraphics[width=\columnwidth]{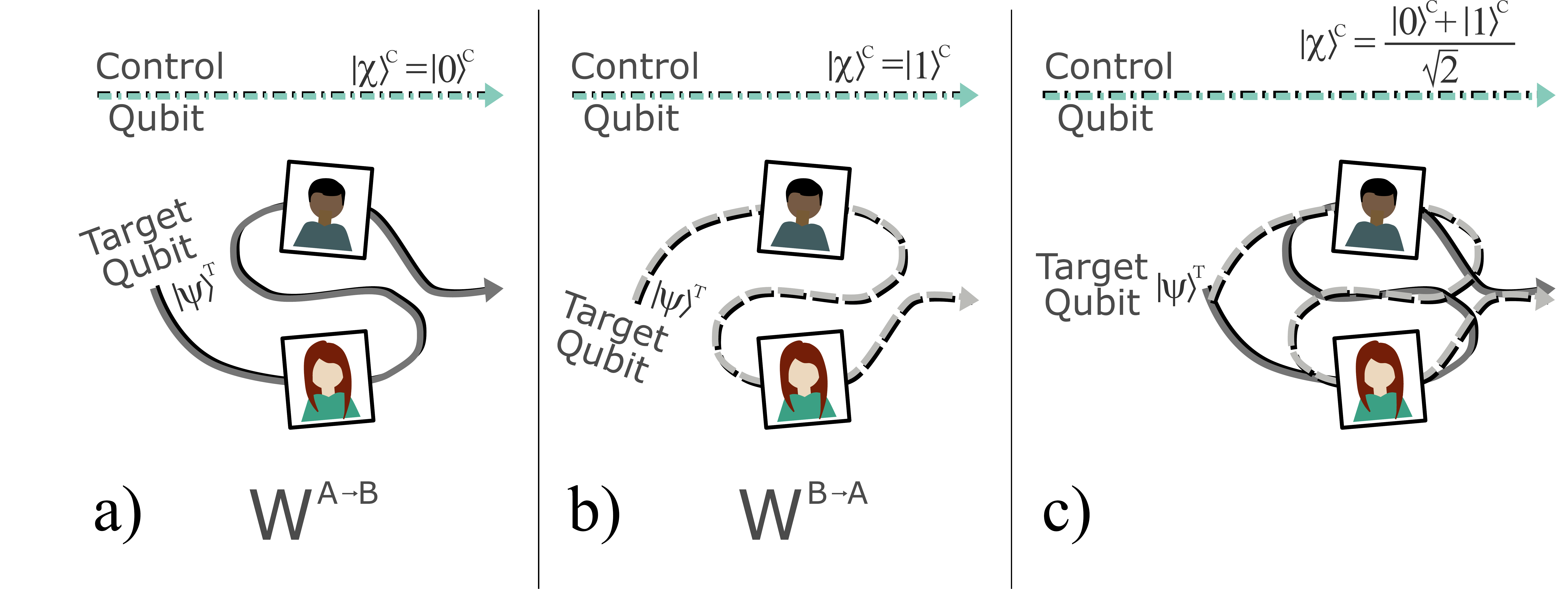}
\captionof{figure}{\footnotesize \textbf{The quantum SWITCH.} Consider a situation wherein the order in which two parties Alice and Bob act on a target qubit $\ket{\psi}^T$ depends on the state of a control qubit in a basis $\left\{ \ket{0},\ket{1}\right\}^C$.  If the control qubit is in the state $\ket{0}^C$ the target qubit is sent first to Alice and then to Bob (Panel \textbf{a)}), while if the control qubit is in the state $\ket{1}^C$, it is sent first to Bob and then to Alice (Panel \textbf{b)}). Both of these situations have a definite causal order, and are described by the process matrices ${W}^{A\rightarrow B}$ and ${W}^{B\rightarrow A}$ (Eq. \ref{eqn:caus_sep}).
If the control qubit is prepared in a superposition state $ \frac{1}{\sqrt{2}}\bigl(\ket{0}+\ket{1}\bigr)^C$, then the entire network is placed into a controlled superposition of being used in the order Alice $\rightarrow$ Bob and in the order Bob $\rightarrow$ Alice (Panel \textbf{c)}). This situation has an indefinite causal order.\vspace{3mm}}
\label{img:quantum_switch}
\end{Figure}

The case described above, called the \textit{quantum SWITCH}, is the first explicit example wherein it was shown that QM does not allow for a well-defined causal order \cite{PhysRevA.88.022318}. The SWITCH was recently experimentally implemented \cite{NatCommun.6} by superimposing the order in which two unitary operations acted. That experiment confirmed that a causally non-ordered quantum circuit can solve a specific computational problem more efficiently than an ordered quantum circuit. But only indirect evidence of indefinite causal order was observed through the demonstration of this computational advantage. The primary goal of our current experiment is, therefore, to provide direct experimental proof of the causal non-separability of the quantum SWITCH.
For this purpose, we used a recently developed theoretical tool: the \textit{causal witness} \cite{NewJournPhys.17.102001}.

\section{Theory}

A causal witness is a carefully designed set of measurements, whose outcome will tell us if a given process is causally ordered or not. An intuitive way to introduce causal witnesses is through the well-known concept of an \textit{entanglement witness} \cite{Gühne20091}. First, recall that a composite quantum system ${\rho}$ lying in a Hilbert space $\mathcal{H}^A \otimes \mathcal{H}^B$ is \textit{separable} or \textit{entangled} depending on whether it can be written in the form ${\rho} = \sum_{i} p_i \rho_i^A \otimes \rho_i^B$ (with $\rho_i^A$ and $\rho_i^B$ states of the subsystems $A$ and $B$ and $0 \leq p_i \leq 1$, $\sum_i p_i =1$) or not. Then it can be shown that for all entangled states ${\rho}^{ent}$ there exists a Hermitian operator ${S}$, called an `entanglement witness', such that $\tr({S}{\rho}^{ent}) < 0$, but $\tr({S}{\rho}^{sep}) \geq 0$ for all separable states ${\rho}^{sep}$. Hence, it follows that if one measures an entanglement witness on a state and finds a negative value the state must be entangled.

A similar quantity was recently introduced to determine if a \textit{process matrix} $W$ is causally separable or not\cite{NatComm.3.012316}. A process matrix (the counterpart of the density matrix in the entanglement witness example) describes causal relations between local laboratories \cite{Veronika}. Consider two observers Alice and Bob who perform local operations $M^A$ and $M^B$ ($M^A$ and $M^B$ can be arbitrary quantum operations, from simple unitary operations to more complex measurement channels).  By local operations we mean that the only connection that Alice and Bob have with the external world is given by the quantum state that they receive from it and the state that they return to it. The process matrix (PM) $W$ then details how this quantum state moves between the two local laboratories (Figure \ref{img:comb}). Hence, it is independent of the individual operations that Alice and Bob perform.
In the case of the quantum SWITCH, the PM first routes the input state to Alice and Bob in superposition, it then connects Alice's output to Bob's input and vice versa, and finally coherently recombines their outputs.

Since a causal witness characterizes a process rather than a state (unlike an entanglement witness), it requires a procedure akin to `process tomography' (\textit{i.e.}, `causal tomography', see \textbf{\textit{Appendix}}, Sec. V).
Namely, we must probe the process with several different input states $\rho^{(\text{in})}$.
Then, for each input state, Alice and Bob implement several different \textit{known} operations, and then we perform a final measurement $D^{(\text{out})}$ (Figure \ref{img:comb}).
In general, $M^A$ and $M^B$ can include measurement operations, so each could have additional measurement outcomes associated with it.
We denote the outcomes of Alice and Bob's local operations by $a$ and $b$, and their choice of operation by $x$ and $y$, respectively.
We label the specific choice of an input state with $z$, and the output of a detection operation with $d$.
With this in mind, the probability of obtaining the outcomes $M_{a, x}^A$, $M_{b, y}^B$, and $D^{(\text{out})}_d$, with the input state $\rho^{(\text{in})}_z$ can be written, using the \textit{Choi-Jamio\l{}kowski isomorphism} (CJ) \cite{CHOI1975285} (see \textbf{\textit{Appendix}}, Sec. VI), as
\begin{align}
\label{eqn:proc_mat}
p\bigl(a, &b, d | x, y, z\bigr) =\\
&= \tr\bigl[\bigl(\rho^{(\text{in})}_z \otimes M_{a, x}^A \otimes M_{b, y}^B \otimes D^{(\text{out})}_d\bigr) \cdot W\bigr]\notag
\end{align}
with $\sum_{a, b, d} p\bigl(a, b, d | x, y, z\bigr) = 1$ for all the possible settings $x$, $y$, $z$ and where $W$ is the PM \cite{NewJournPhys.17.102001}.

\begin{Figure}
\centering
\includegraphics[width=.8\columnwidth]{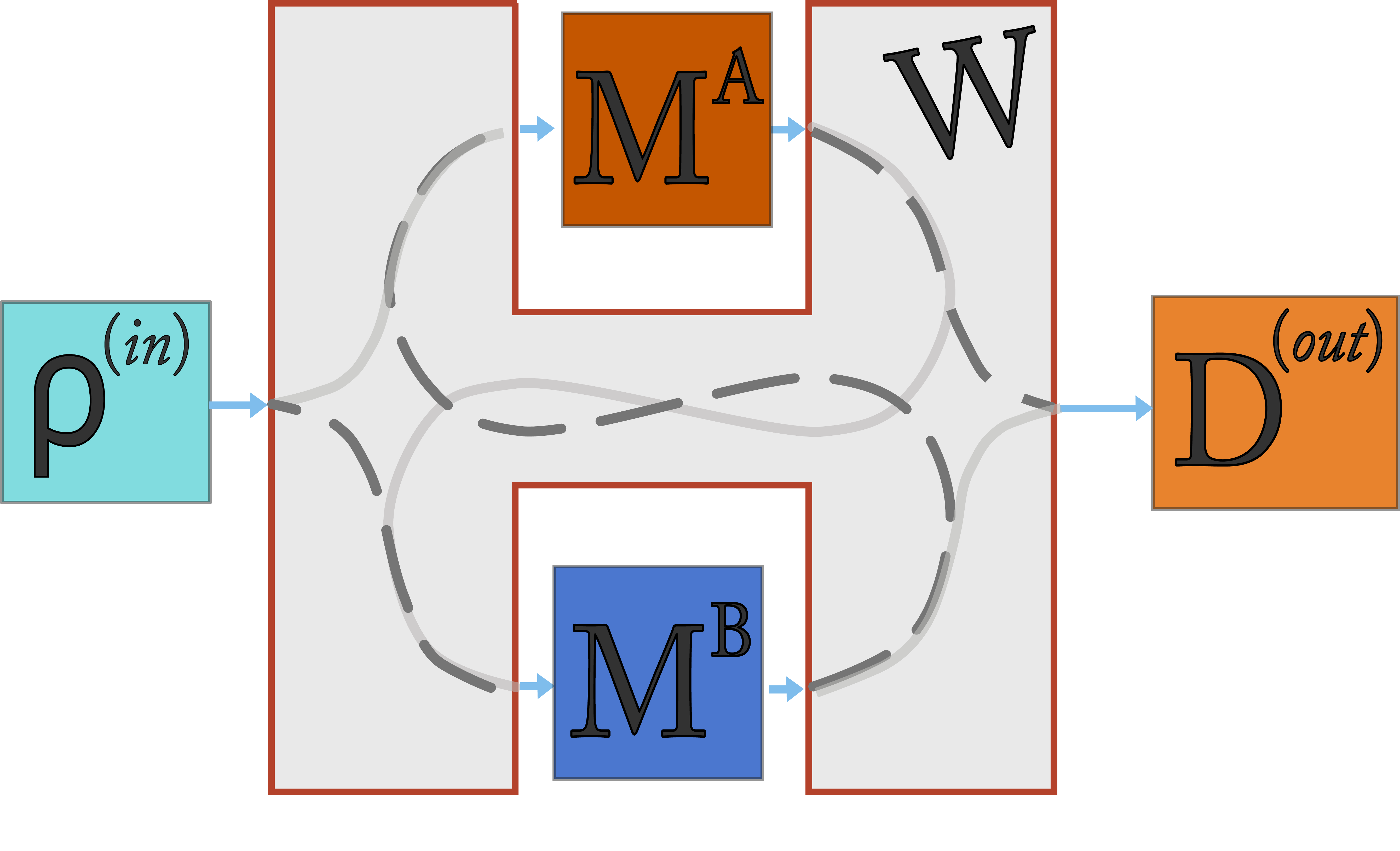}
\captionof{figure}{\footnotesize \textbf{A process matrix representation of Figure \ref{img:quantum_switch}.}
The process matrix $W$ describes the ``links'' between Alice and Bob.
For example, it could simply route the input state $\rho^{(\text{in})}$ to Alice $M^A$ and then to Bob $M^B$ (solid line), or vice versa (dashed line).
In the case of the quantum SWITCH, it creates a superposition of these two paths, conditioned on the state of a control qubit.
The input state $\rho^{(\text{in})}$, the two local operations $M^A$ and $M^B$, and the final measurement $D^{(\text{out})}$ must all be controllable and known a-priori.
The unknown process is represented by the process matrix (shaded grey area labelled $W$).
A causal witness quantifies the causal non-separability of $W$.\vspace{3mm}
}
\label{img:comb}
\end{Figure}

To calculate these probabilities for the quantum SWITCH we must construct its PM, which we will call $W_{\text{SWITCH}}$. To do this we will again use the CJ isomorphism. 
As a first step, consider the identity channel from the output space $\mathcal{H}^{P_1}_{\text{out}}$ of a party $P_1$ to the input space $\mathcal{H}^{P_2}_{\text{in}}$ of a second party $P_2$.  To describe this as a process matrix we can write it as a projector onto a process vector in the `{double-ket notation}' \cite{PhysRevA.43.44, PhysRevLett.84.3486}:
\begin{equation}\label{eq:simplePM}
\ket{\mathbb{1}\rangle}^{\mathcal{H}^X_{\text{in/out}} \mathcal{H}^Y_{\text{in/out}}} = \sum_j \ket{j}^{\mathcal{H}^X_{\text{in/out}}}\otimes\ket{j}^{\mathcal{H}^Y_{\text{in/out}}},
\end{equation}
where $j$ labels a basis over the spaces.
We can now use this PM to describe an input state passing first to Alice ($\mathcal{H}^{A}_{\text{in}}\rightarrow\mathcal{H}^{A}_{\text{out}}$), then to Bob ($\mathcal{H}^{B}_{\text{in}}\rightarrow\mathcal{H}^{B}_{\text{out}}$), and finally to the output space ($\in \mathcal{H}^{\text{(out)}}$). This process is described by
\begin{equation}\label{eq:wVEC}
 \ket{{w}^{A\rightarrow B}} = \ket{\mathbb{1}\rangle}^{\mathcal{H}^{\text{(in)}} \mathcal{H}^A_{\text{in}}}\ket{\mathbb{1}\rangle}^{\mathcal{H}^A_{\text{out}} \mathcal{H}^B_{\text{in}}}\ket{\mathbb{1}\rangle}^{\mathcal{H}^B_{\text{out}} \mathcal{H}^{\text{(out)}}}.
\end{equation}
Alice and Bob are free to perform measurements $M^A: \mathcal{H}^{A}_{\text{in}}\rightarrow\mathcal{H}^{A}_{\text{out}}$ and $M^B: \mathcal{H}^{B}_{\text{in}}\rightarrow\mathcal{H}^{B}_{\text{out}}$, respectively, but they are not part of the above process vector.
Note that swapping the order of Alice and Bob is as simple as swapping the labels $A$ and $B$. The vectors $\ket{{w}^{A\rightarrow B}}$ (describing `Alice acts before Bob') and $\ket{{w}^{B\rightarrow A}}$ (describing `Bob acts before Alice') both have a well-defined causal order (Figure \ref{img:quantum_switch}, Panels \textbf{a)} and \textbf{b)}). 

We are now in the position to construct the PM of the quantum SWITCH.
Recall that for the quantum SWITCH the control qubit's state sets the relative amplitudes of Alice $\rightarrow$ Bob and Bob $\rightarrow$ Alice. Thus the process vector of the quantum SWITCH (when the control qubit initially in the state $\frac{\ket{0}^C+\ket{1}^C}{\sqrt{2}}$) is quite simply:
\begin{equation}
\ket{w_\text{SWITCH}} = \frac{1}{\sqrt{2}}\Bigl(\ket{{w}^{A\rightarrow B}}\ket{0}^{C} + \ket{{w}^{B\rightarrow A}}\ket{1}^{C}\Bigr).
\end{equation}
For the causal witness we will consider here, we will only measure the state of the control qubit after the SWITCH.
Thus, we need to construct the PM taking an input state, and returning the state of the control qubit.
This is done by tracing over the SWITCH output (\textit{i.e.}, the target qubit) and fixing the state of the control qubit to be $\ket{0} + \ket{1}$.
So the PM to compute the final state of the control qubit is represented by the PM:
\begin{equation}
W_{\text{SWITCH}} = \tr_{\mathcal{H}^{\text{(out)}}}\bigl(\ket{w_\text{SWITCH}}\bra{w_\text{SWITCH}}\bigr),
\label{eqn:switch}
\end{equation}
where $\tr_{\mathcal{H}^{\text{(out)}}}(\cdot)$ is the partial trace over the output system qubit.

Using the same formalism, one can also concisely describe all causally separable processes. 
Consider two general PMs linking the two local laboratories $A$ and $B$, ${W}^{A\rightarrow B}$ and ${W}^{B\rightarrow A}$.
Here, contrary to Eq. \ref{eq:wVEC}, the link between the laboratories is no longer the identity channel.
Then by simply taking an incoherent mixture of two, one can describe all possible causally separable processes \cite{NewJournPhys.17.102001}:
\begin{equation}
\label{eqn:caus_sep}
{W}^{sep} := p {W}^{A\rightarrow B} + (1-p){W}^{B\rightarrow A},
\end{equation}
where $0\leq p \leq 1$. Physically, this can be understood as each run of the process having a well-defined order, with Alice acting first with probability $p$ and Bob acting first with probability $1-p$.

Causal witnesses are designed to distinguish between causally separable (Eq. \ref{eqn:caus_sep}) and causally non-separable process matrices (such as Eq. \ref{eqn:switch}). Indeed, it has been proven in \cite{NewJournPhys.17.102001} that, analogously to the entanglement case, for all causally non-separable process matrices ${W}^{n-sep}$, there exists a Hermitian operator ${S}$, called a causal witness, such that
\begin{equation}
\tr({S} {W}^{n-sep}) < 0,
\end{equation}
but $\tr({S} {W}^{sep}) \geq 0$ for all causally separable process matrices ${W}^{sep}$.
Using the notation defined above (Eq. \ref{eqn:proc_mat}), a causal witness $S$ can be explicitly written as
\begin{equation}
\label{eqn:S}
S = \sum_{\substack{a,b,d\\x,y,z}} \alpha_{a,b,d,x,y,z} \cdot \rho^{(\text{in})}_z \otimes M_{a, x}^A \otimes M_{b, y}^B \otimes D^{(\text{out})}_d.
\end{equation}
There are six sums, one over the input state (label $z$), one over each of Alice and Bob's operation choices (labels $x$ and $y$) and outcomes (labels $a$ and $b$), and one over the detection operation outputs (label $d$). Each term is weighted by a real coefficient $\alpha_{a,b,d,x,y,z}$, whose value verify a normalization condition on $S$. 
The choices of the various $\alpha_{a,b,d,x,y,z}$, $\rho^{(\text{in})}_z$, $M_{k_1, k_2}^X$, and $D^{(\text{out})}_d$ define a specific witness.
Numerical algorithms exist to find the optimal causal witness to test for a specific causally non-separable process (see \textbf{\textit{Appendix}}, Sec. V).

The quantity $-\tr({S} {W})$ is an analogue of the so-called `generalized robustness of entanglement' \cite{PhysRevA.67.054305}, which quantifies the resistance of the entanglement to noise. Similarly, $-\tr({S} {W})$ corresponds to the maximum `amount of noise that the process under examination can tolerate while remaining causally non-separable \cite{NewJournPhys.17.102001} (below we will define this `amount of noise').
Hence, we will refer to this quantity as the `causal non-separability' (CNS) of a process ${W}$:
\begin{equation}
\text{CNS}({W}) := -\tr\bigl({S} {{W}}\bigr).
\end{equation}
When the $-\tr\bigl({S} {{W}}\bigr)<0$, we define the $\text{CNS}({W})$ to be zero.
A given process can be proven to be causally non-separable by having a positive value of the CNS. The larger its value, the further away the process is from a causally separable process (\textit{i.e.}, it can tolerate more noise before becoming causally separable).

In addition to the causal non-separable nature of ${W}$, the magnitude of the experimentally accessible value of the CNS depends on the local operations that Alice and Bob implement. In order to fully assess the CNS, Alice and Bob must be able to implement a complete set of operators.
The experimentally certifiable CNS (hereafter referred to as $\text{CNS}_{\text{exp}}({W}) = -\tr\bigl({S}_{\text{exp}} {{W}}\bigr)$) depends on the amount of information one can gather about the process ${W}$. The more information one acquires, the more noise can be tolerated while still being able to confirm that the process is causally non-separable.
To achieve $\text{CNS}_{\text{exp}}({W})=\text{CNS}({W})$ one must, in general, perform full causal tomography (see \textbf{\textit{Appendix}}, Sec. V).

Since a unitary operation cannot extract any explicit information from the manipulated state, neither Alice nor Bob can gain any knowledge about their received state when applying only such gates. 
Thus, with only unitary operations the maximal value of $\text{CNS}_{\text{exp}}({W})$ is limited.
 However, if the unitary operations are replaced with projective measurements then, roughly speaking, information about the process at different points throughout the SWITCH can be extracted.
In fact, if Alice and Bob both have measure and reprepare (MR) operations, one can achieve $\text{CNS}_{\text{exp}}({W})=\text{CNS}({W})$.
Remarkably, it turns out that giving one party a MR operation and the other a unitary operation still increases $\text{CNS}_{\text{exp}}({W})$ substantially.
Because of the experimental challenges of coherently adding MR operations, in our experiment Alice performs an MR operation and Bob implements a unitary channel.
{So the causal witness we will measure depends both on Alice's outcome (performed inside the SWITCH) and our final measurement outcome.}

\section{Experiment}

To experimentally implement the quantum SWITCH, we need a control and a target qubit. In our experiment we encode a control qubit in a path degree of freedom (DOF) of a photon and a target qubit in the same photon's polarization. 
This is convenient as Bob's unitary gate can be implemented easily with three waveplates (WPs), while Alice can perform a projective measurement with WPs and a polarizing beamsplitter (PBS). Note that other schemes for implementing the SWITCH have been also proposed \cite{PhysRevA.89.030303,PhysRevA.93.052321}.

Although the realization of the unitary channel is straightforward, a short remark is necessary concerning Alice's measurement.
It is clear that a PBS enables one to distinguish the polarization of an incoming photon.
However, a PBS gives rise to additional spatial modes (\textit{i.e.}, after the PBS there are two output paths). 
These two spatial modes can be considered as a new spatial qubit. 
Then, the action of the PBS is to couple the polarization qubit to this additional qubit. 
This is formally equivalent to a \textit{von Neumann system-probe coupling}, which can model any projective measurement \cite{vonneumann}.
In our experiment, the polarization qubit is the system, and it is coupled (via the PBS) to an additional spatial qubit which is the probe.
We can read out information about the system by measuring the probe (with a photon detector) at a later time.
This solves the non-trivial problem of realizing a measurement operation \textit{inside} a quantum SWITCH.
Most approaches to acquire information inside the SWITCH would lead to distinguishing information about the order in which the operations were applied, destroying the quantum superposition.
In our solution, however, since the probe qubit is not measured until the information about the order of application of the operations is erased, the entire process can remain coherent. 
This solution also works deterministically, \textit{i.e.}, both of Alice's outcomes are retained.  It also allows Alice to implement a measurement-dependent repreparation by placing different WPs in each of the two outcome modes.

Our implementation of the quantum SWITCH draws inspiration from a previous experiment \cite{NatCommun.6}, in which only orders of unitary operations were superimposed. Therefore, as in \cite{NatCommun.6}, our experimental skeleton consists of a Mach-Zehnder interferometer (MZI) with a loop in each arm. However, because Alice's MR channel adds an additional path degree of freedom, we need an extra interferometric loop.

\begin{Figure}
\centering
\includegraphics[width=\columnwidth]{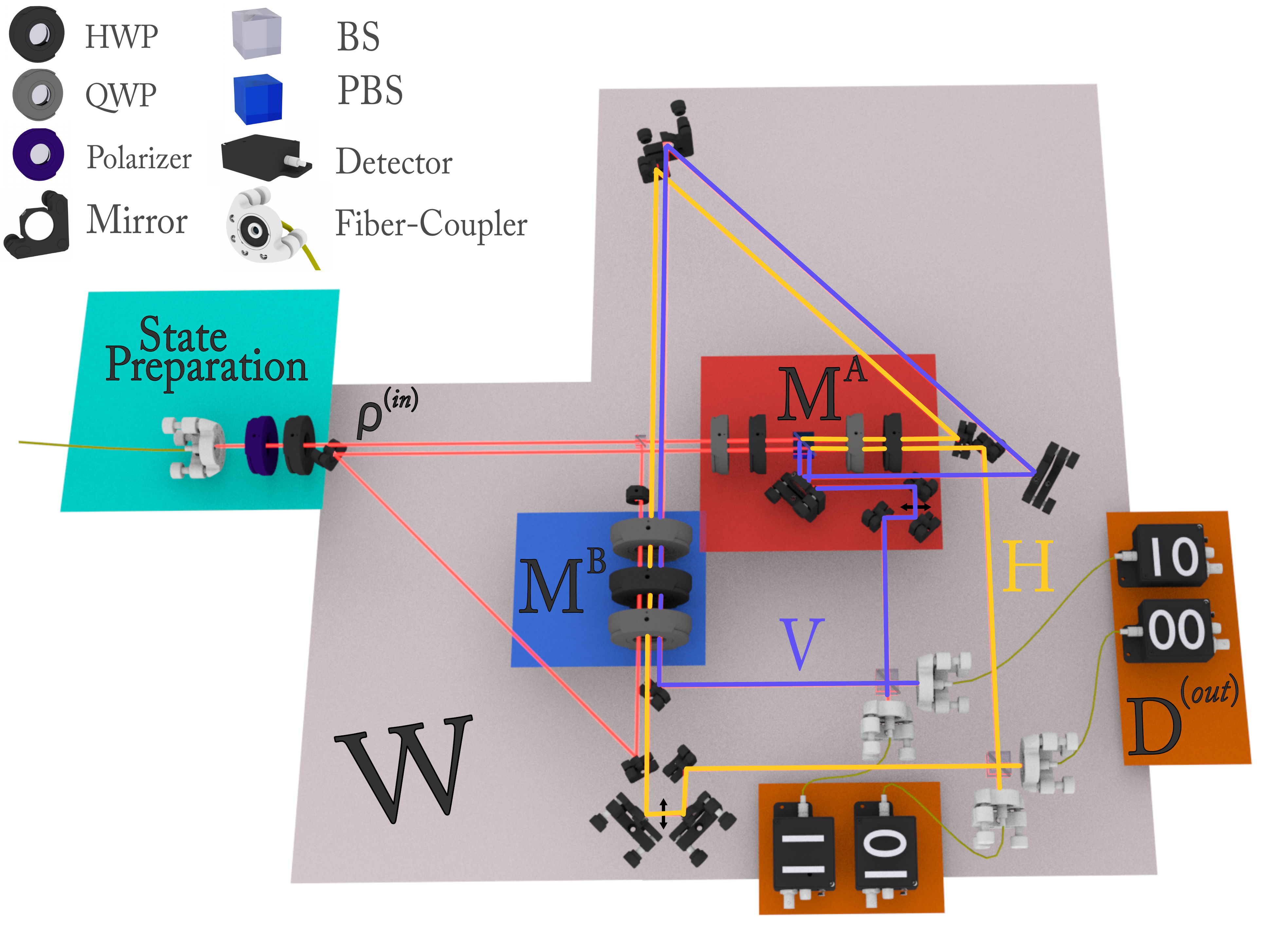}
\captionof{figure}{\footnotesize \textbf{Experimental Set-Up.} A sketch of our experiment to verify the causal non-separability of the quantum SWITCH. We produce pairs of single photons using a Type-II spontaneous-parametric-down-conversion source (not shown here). One of the photons is used as trigger, one is sent to the experiment. The experiment body consists of two Mach-Zehnder interferometers, with loops in their arms. The qubit control,  encoded in a path degree of freedom, dictates the order in which the operations $M^A$ and $M^B$ are applied to the target qubit (encoded in the same photon's polarization)
Alice implements $M^A$ (a measurement and repreparation), and Bob implements $M^B$ (a unitary operation). 
After the interferometer the control qubit is measured, \textit{i.e.}, we check if the photon exits port 1 or port 2.
The detector labels refer to Alice's measurement outcome (first digit) and the final measurement outcome (second digit).
When Alice measures $0$ (1) the photon exits via the external ---yellow--- (internal ---blue) interferometer.
When the final measurement result is 0 (1) the photon exits through a horizontally (vertically) drawn port.
A half waveplate at $0$\degree$\,$ was used in the reflected arm of the first BS in order to compensate the acquired additional phase. Acronyms in the figure are defined as follows: QWP, quarter waveplate; HWP, half waveplate; BS, beam splitter; PBS, polarizer beam splitter.\vspace{2mm}}
\label{img:experimental_apparatus1}
\end{Figure}

A sketch of our experimental apparatus is presented in Figure \ref{img:experimental_apparatus1}.  The first step is to set the state of the system qubit (encoded in the polarization) with a polarizer and a HWP. Then the photon impinges on a $50/50$ beam splitter (BS); this sets the state of the control qubit (encoded in a path DOF) in $\ket{+}$. Depending on the state of the path qubit, the photon is sent to either Alice (who performs $M^A$) and then Bob (who performs $U^B$), or vice versa. As described above, $M^A$ is a projective measurement (a sequence of two WPs and a PBS) and a corresponding repreparation (a sequence of two WPs in only one of the PBS outputs), and $U^B$ is a unitary gate (a sequence of three WPs). 
Since the PBS adds a second path qubit this results in four path modes, encoding both the state of the control qubit and the outcome of the MR channel.
Referring to Figure \ref{img:experimental_apparatus1}, the {external} (yellow) interferometer arises from the outcome $H$ - also referred to as a logical $0$ - and the {internal} (blue) one from the outcome $V$ - a logical $1$. We finalize the SWITCH by erasing the information about the order of the gates.
This can be done by applying a Hadamard gate to the control qubit. Since the control qubit is a path qubit a Hadamard gate can be implemented with a $50/50$ BS. However, in our experiment there are two path-qubits (the control qubit and Alice's ancilla measurement qubit).
Thus, we must use two $50/50$ BSs: one BS to interfere the control qubit when Alice's ancilla qubit is in the state $\ket{0}$, and one BS when it is in the state $\ket{1}$.
Finally, each of the four outputs are coupled into single-mode fibers, which are each connected to single-photon detectors (SPD). Then detecting a photon in one of the four modes yields the result of both the measurement of the control qubit in the superposition basis and Alice's measurement (see the detector labels in Figure \ref{img:experimental_apparatus1}).

We wish to evaluate the CNS of our quantum SWITCH by experimentally estimating the expectation value of a causal witness $S$ (Eq. \ref{eqn:S}). In other words, we want to assess $\tr(S_{\text{exp}} W_{\text{SWITCH}})$, where here $W_{\text{SWITCH}}$ refers to the PM describing our experiment.
Since the trace is linear, this can be done by implementing one term in the sum of $S$ (Eq. \ref{eqn:S}) at a time.
To estimate a single term, an input state is injected into the SWITCH, Alice and Bob each perform an operation inside, and then the outputs of the overall process are measured.
Since the control qubit measurement and Alice's measurement are both single-qubit projective measurements, there are a total of four possible outcomes.
For each measurement setting, different input states are sent into the SWITCH and the probabilities of each outcome are experimentally estimated by sending multiple copies of the same input state.
To compute the final value of the $\text{CNS}_{\text{exp}}(W_{\text{SWITCH}})$, the results of these measurements are weighted by the corresponding $\alpha_{a,b,d,x,y,z}$ and summed.

The number of terms in the sum of Eq. \ref{eqn:S} is determined by the specific witness we wish to evaluate.
In general, Alice and Bob must each implement a set of operators forming a basis over their channels.
For Bob's unitary channel this requires $10$ elements and for Alice's MR channel $16$ \cite{NewJournPhys.17.102001}.
In our case, we formed  Alice's basis with \textit{four} (non-commutative) projection operators, and \textit{three} unitary repreparation operators when the outcome was $H$ and \textit{one} operator (the identity operator) when the outcome was $V$. This corresponds to $12$ MR channels when the outcome of Alice's measurement is $H$ and $4$ when it is $V$, for a total of $16$ MR operators. For Bob we implement all $10$ unitaries.

Varying the input state can make $\text{CNS}_{\text{exp}}(W_{\text{SWITCH}})$ more robust to noise. Hence, for our experiment we used \textit{three} different input states: $\ket{H}$, $\ket{V}$ and $\ket{+}$.
Finally, we implemented \textit{two} different measurement operators $D^{\text{(out)}}$ on the control qubit (corresponding to the two outcomes of the projection onto basis $\left\lbrace\ket{\pm}=\frac{\ket{0}\pm\ket{1}}{\sqrt{2}}\right\rbrace$).
Thus, for our experiment, the calculation of $\text{CNS}_{\text{exp}}(W_{\text{SWITCH}})$ translates into
\begin{align}
\label{eqn:CNS}
&\text{CNS}_{\text{exp}}(W_{\text{SWITCH}}) =\\
&= -\sum\limits_{z=0}^2 \sum\limits_{a=0}^{1}\sum\limits_{x=0}^{11}\sum\limits_{y=0}^{9} \sum\limits_{d=0}^1 \alpha_{a,d,x,y,z} \; p\bigl(a, d | x, y, z\bigr),\notag
\end{align}
here we do not need the sum over $b$, since Bob's unitaries do not have an outcome. The probability in Eq. \ref{eqn:CNS} is defined as
\begin{align}
&p\bigl(a, d | x, y, z\bigr) :=\\
& :=\tr\bigl[\bigl(\rho^{(\text{in})}_z \otimes M_{a, x}^A \otimes U_{y}^B \otimes D^{(\text{out})}_d\bigr) \cdot W_{\text{SWITCH}} \bigr].\notag
\end{align}
We must experimentally estimate all of these probabilities to evaluate $\text{CNS}_{\text{exp}}(W_{\text{SWITCH}})$.
There are $1440$ terms in this sum. However, four outcomes (two from of Alice's measurement and two from the final detection) are collected simultaneously (experimentally, this means the counts of four SPDs are collected in one setting). Therefore, we need $360$ different experimental settings. However, for our witness, of the $360$ pre-factors $\alpha_{a,d,x,y,z}$, $101$ are equal to zero, so there are actually only $259$ relevant experimental settings.

With this in place, we are able to experimentally measure the $\text{CNS}_{\text{exp}}(W_{\text{SWITCH}})$ (for information relating experimental visibility, stability and data taking procedure, see \textbf{\textit{Appendix}}, Sec. II-IV). Figure \ref{img:first_70} shows some of the probabilities $p\bigl(a, d | x, y, z\bigr)$ (Eq. \ref{eqn:CNS}) for the four outcomes, \textit{i.e.}, for Alice $a=0,1$ and our final measurement $d=0,1$ (the remainder are shown in the \textbf{\textit{Appendix}}).
In Figure \ref{img:first_70}, the experimentally obtained values are the blue dots and the theoretical predictions are the bars. 

\begin{Figure}
\centering
\includegraphics[width=\columnwidth]{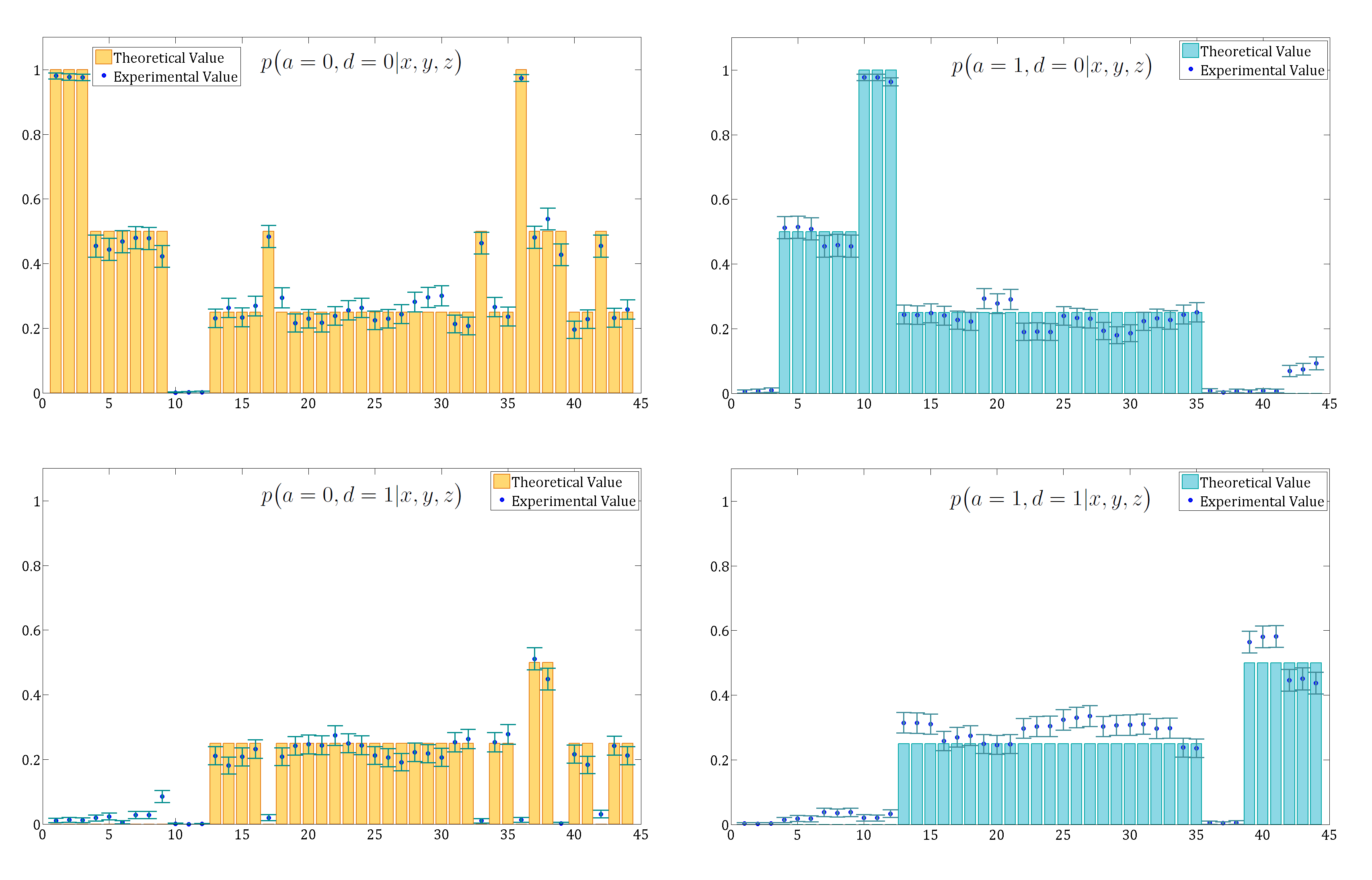}
\captionof{figure}{\footnotesize \textbf{Experimentally estimated probabilities.} Each data point represents a probability $p\bigl(a, d | x, y, z\bigr)$ in Eq. \ref{eqn:CNS} for $a = 0, 1$ and $d = 0, 1$. The blue dots represent the experimental result and the bars the theoretical prediction. The yellow (blue) bars refer to the external (internal) interferometer. The $x$-axis is the measurement number. \vspace{3mm}}
\label{img:first_70}
\end{Figure}

Our main source of error is phase fluctuations in the two interferometers. We therefore performed a separate measurement (presented in the \textbf{\textit{Appendix}}, Sec. IV) to characterize this error.
The error bars in Figure \ref{img:first_70} represent both these phase errors and Poissonian errors due to finite counts.
These errors do not take into account systematic errors, such as waveplate miscalibration, since such systematic errors represent a deviation of our experimental SWITCH from the ideal SWITCH.

We can now obtain a value for the CNS of our process by weighting the data presented in Figure \ref{img:first_70} (and Supplementary Figures 1-3) by $\alpha_{a,d,x,y,z}$ and then summing them.  The result is
\begin{equation}
\text{CNS}_{\text{exp}}(W_{\text{SWITCH}}) = 0.202\pm 0.029.
\end{equation}
The error bar on $\text{CNS}_{\text{exp}}(W_{\text{SWITCH}})$ was calculated by Gaussian error propagation from the errors of the individual probabilities.
The theoretical maximum value for $\text{CNS}_{\text{exp}}(W_{\text{SWITCH}})$ is $0.2842$.  The disagreement between this and our measured result is caused primarily by two effects.  First, given the reduced visibility of the interferometers (which we will discuss in detail shortly) the maximal value for $\text{CNS}_{\text{exp}}(W_{\text{SWITCH}})$ is $0.2523$, when the visibility is $0.9539$.  The remaining discrepancy comes from systematic errors, such as waveplate miscalibration, which effectively mean that the unitaries Alice and Bob implement differ slightly from their targets.  For example, we estimate, using a simple Monte Carlo simulation, that a waveplate calibration error of $3^\circ$ would explain this discrepancy, leading to a drop in the CNS of approximately $0.043$.
Still, given our measured result, we can conclude that our process is causally non-separable by a margin of approximately seven standard deviations. This large margin demonstrates the effectiveness of performing a measurement operation inside the quantum SWITCH.

As mentioned above, the causal non-separability (as measured using a causal witness) can be considered a measurement of how much noise can be added to the process before it becomes causally separable. 
The $\text{{CNS}}_{\text{{exp}}}$ we have discussed so far refers to a `worst-case noise' model \cite{NewJournPhys.17.102001}, wherein the desired process is replaced with the process that can do most damage to its causal nonseparability with a probability
\begin{equation}
p_{\text{worst-case}} := \dfrac{\text{CNS}_{\text{exp}}(W_{\text{SWITCH}})}{1+\text{CNS}_{\text{exp}}(W_{\text{SWITCH}})}.
\end{equation} 
Since the replacement is done with the worst-case process, this is a a
lower bound on the ‘probability of noise’ that can be tolerated (see \textbf{\textit{Appendix}}, Sec. V).
For our process $p_{\text{worst-case}}=0.168 \pm 0.001$.

We studied the effect of the noise most relevant to our experiment. 
Namely, dephasing the control qubit but not the system qubit.  
This noise is the strongest in our setup because the control qubit is encoded in a path DOF, which must remain interferometrically stable (see Ref. \cite{ScienRep.6.26018} for the formal definition of this noise model). 
We realized this noise by imbalancing the path length of the interferometers by more than the photons' coherence length.
The experimental signature of this imbalance is a reduced visibility of the interferometer, we measured the CNS for several visibilities between $0.95$ and $0.06$.
Figure \ref{img:CWvaryingVis} shows a decrease in the expectation value of $-\tr\bigl({S} {W_{\text{SWITCH}}}\bigr)$ as the noise increases. 
There is an offset between the experimental data and the the theoretical prediction due to systematic errors. 
However, both theory and experiment follow the same trend.
By extrapolating our fit of the experimental data to $-\tr\bigl({S} {W_{\text{SWITCH}}}\bigr)=0$ (where the process becomes causally separable), we observe a `noise tolerance' of $0.342$ for this type of noise. 
As expected this is larger than our experimentally measured $p_{\text{worst-case}}$, indicating that it is indeed a lower bound.

\begin{Figure}
\centering
\includegraphics[width=\columnwidth]{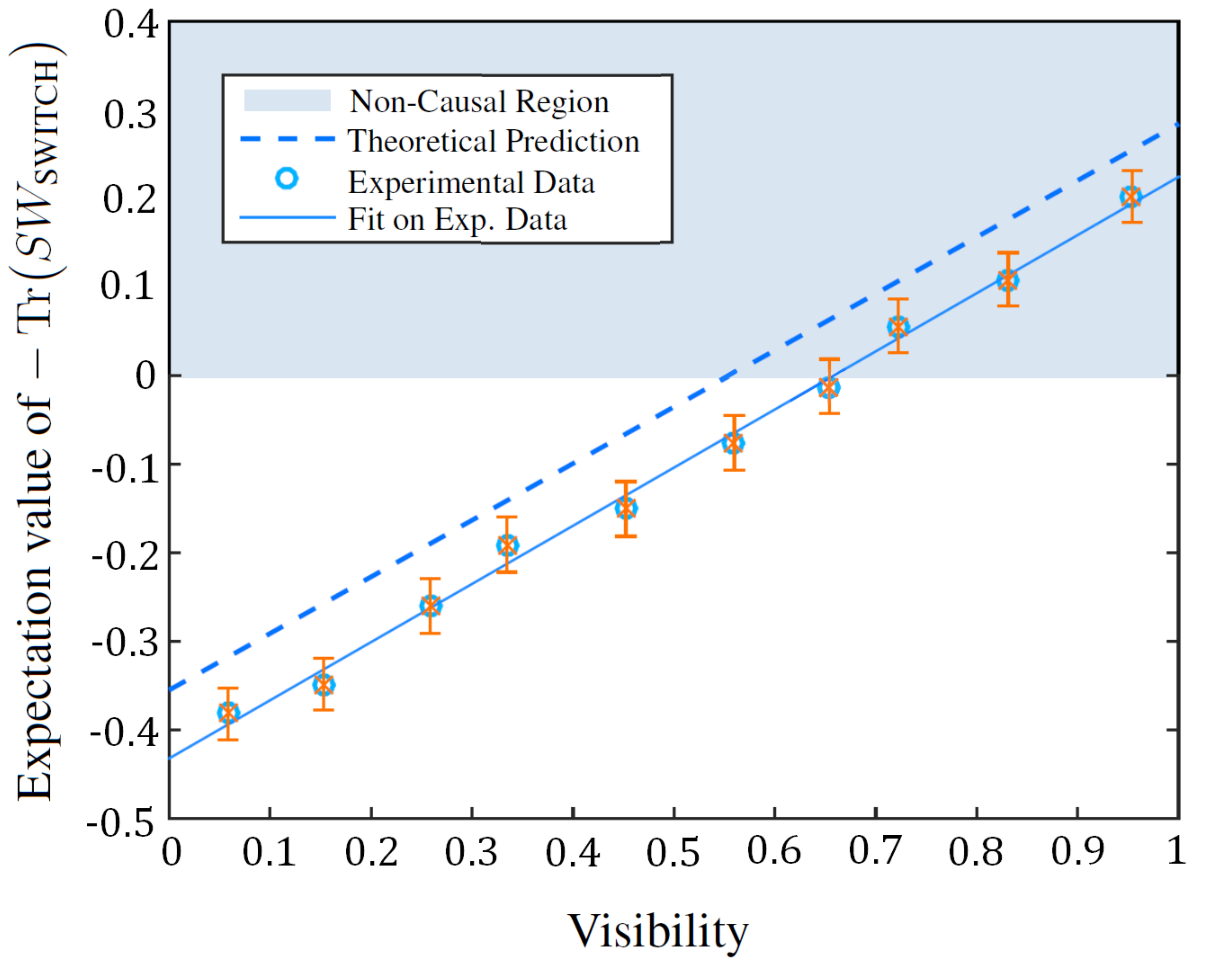}
\captionof{figure}{\footnotesize \textbf{Expectation value of the causal witness (
$-\tr\bigl({S} {W_{\text{SWITCH}}}\bigr)$) in the presence of noise.} As the control qubit (intitially in $\ket{+}$) is decohered, the superposition of causal orders becomes an incoherent mixture of causal orders.
Hence, the causal non-separability of the SWITCH is gradually lost. 
The plot shows the causal non-separability of our experimentally implemented SWITCH as the visibility of the two interferometers is decreased (from right to left). 
The experimental data linearly decreases with visibility just as theory (dashed line) predicts. 
The gap between theory and experiment is because of systematic errors.
The visibility ($x$ axis) is a measure of the dephasing strength on the control qubit.}
\label{img:CWvaryingVis}
\end{Figure}

\section{Discussion}
Our experiment demonstrates how to perform a measurement \textit{inside} a quantum SWITCH without destroying the superposition of causal orders. The task was only assumed to be possible in Ref. \cite{NewJournPhys.17.102001}, but no method to accomplish it was proposed. The difficulty is that performing a standard measurement necessarily reveals the time at which it is performed, and thus whether it is performed before or after the partner's operation. Consequently, the superposition of causal orders becomes incoherent. Our way around this is to break the measurement into two steps: first the system coherently interacts with an ancilla through a unitary operation (namely, the additional path modes introduced by the local operation in our experiment). Then, after finalizing the quantum SWITCH (interfering these modes), the ancilla is measured. This allows us to make a ``coherent measurement'' at different times and then erase the ordering information. 

We demonstrated the causal non-separability of our experimental apparatus by measuring a causal witness. With the ability to perform a measurement inside the SWITCH we were able to increase the robustness of the causal witness to noise.
Previous experimental work only indirectly accessed the causal non-separability of the SWITCH and, moreover, only used unitary gates in the SWITCH\cite{NatCommun.6}. Although some other experiments \cite{NatPhys.11.414, MacLean, white} have also studied the topic of causal relations in quantum mechanics, they focused on a different aspect.  For example, in \cite{NatPhys.11.414, MacLean} they distinguish between different causal structures, rather than creating a genuinely indefinite causal order, as in our work.
In fact, the incoherent mixture \cite{NatPhys.11.414} and a quantum superposition \cite{MacLean} of different causal relations reported previously are both compatible with one party in the past and the other in the future. Thus, in our language they correspond to causally-ordered processes.

Our work represents the first experimental realization of a quantum superposition of orders of non-unitary channels, and the first measurement of a causal witness. 
We believe this will be an important step towards the realization of quantum superpositions of the order of more elaborate processes. 
Since it has been theoretically demonstrated that causally non-ordered processes can give rise to a reduction in the query complexity of certain tasks \cite{Hardy2009,PhysRevA.86.040301,PhysRevLett.113.250402}, and lead to more efficient communication channels \cite{PhysRevA.92.052326,Philippe}, it is important to study new techniques to create more complex causally non-ordered processes.
In fact, we already see an advantage in our current work. Making a measurement inside the quantum SWITCH made our experiment more robust to noise and allowed us to demonstrate, by approximately seven standard deviations, that our set-up cannot be described by a causally ordered process.

\vspace{2mm}
\paragraph{Acknowledgements} We thank I. Alonso Calafell for assisting with the electronics and F. Costa, F. Massa, M. Zych, and C. Branciard for useful discussions. We acknowledge support from the European Commission, EQUAM (No. 323714), PICQUE (No. 608062), GRASP (No.613024), QUCHIP (No.641039), and RAQUEL (No. 323970); the Austrian Science Fund (FWF) through START (Y585-N20), the doctoral programme CoQuS, and Individual Project (No. 2462); the United States Air Force Office of Scientific Research (FA9550-16-1-0004); the John Templeton Foundation; and the Foundational Questions Institute (FQXi).

\end{multicols}
\bibliography{QDC_BIB}
\bibliographystyle{mateus}

\newpage
\paragraph{Appendix I. Single Photon Source}\vspace{3mm}
We generate heralded single-photons using a Type-II spontaneous parametric down-conversion (SPDC) process in a Sagnac loop \cite{PhysRevA.73.012316}. The Sagnac loop is realized using a dual-wavelength PBS and two mirrors. The SPDC crystal is a 20-mm-long \textit{periodically-poled crystal Potassium Titanyl Phosphate} (PPKTP) crystal. The crystal is pumped by a 23.7 mW diode laser centred at 395 nm. The polarization of the laser is set to be horizontal. With this, we generate degenerate pairs of single photons centred at 790 nm, in a separable polarization state $\ket{H}\ket{V}$. Polarizers in the signal and idler modes are used to ensure that the polarization is in a  well-defined state. The down-converted photons are coupled into single-mode fibers. One photon is sent directly to an single-photon detector, and is used to herald the other photon's presence for the experiment, while the other is sent to our experiment.
After passing though the experiment we observe a coincidence rate between the herald detector and the four final-measurement detectors of 3750 pairs per second.

\paragraph{Appendix II. Implementing Alice and Bob's Channels}\vspace{3mm}

As discussed in the main text, in order to experimentally measure a causal witness Alice and Bob need to implement a series of quantum channels on a polarization qubit inside the quantum SWITCH.  Alice must perform a measure and reprepare (MR) channel, while Bob implements a unitary channel. 
Alice measures in four different bases. 
We define her different bases by a unitary operator preceding a projective measurement in the basis $\left\lbrace \ket{0},\ket{1}\right\rbrace$.
Alice's pre-measurement operators are listed in the first column of Table \ref{tab:op}.
When her outcome is $\ket{0}$ (in a given basis), Alice implements one of three different repreparation operators (second column of Table \ref{tab:op}).
On the other hand, when her outcome is $\ket{1}$ she performs the identity channel. Thus she has $16$ different measure and reprepare (MR) maps. 
Bob simply implements $10$ different unitary operators (third column of Table \ref{tab:op}).
In Table \ref{tab:op}, the $\sigma_i$ operators are the Pauli operators and $\mathcal{H}$ is the Hadamard gate. 

\begin{center}
\begin{table}[htb]
\caption{\footnotesize \textbf{List of operators performed by the two parties.} The table shows Alice's four measurement operators, and her three repreparation operators which she applies when her outcome is $\ket{0}$; when her outcome is $\ket{1}$ she performs the identity. Bob's ten unitary operators are shown in column three.}
\label{tab:op}
\centering
\begin{tabular}{l|l|l}
\toprule
$\quad${\textbf{Alice - Meas.}} & $\quad${\textbf{Alice - Reprep.}} & $\qquad\;\qquad${\textbf{Bob - Unit. Op.}}\\
\midrule
${\gra (1)}\quad\quad$ $\begin{pmatrix} 
1 & 0\\
0 & i
\end{pmatrix}$ &
 ${\gra (1)}\quad\quad\;\;\;\,$  $\begin{pmatrix} 
1 & 0\\
0 & i
\end{pmatrix}$ &
 $\,{\gra (1)}\quad\;$  $\begin{pmatrix} 
1 & 0\\
0 & 1
\end{pmatrix}$ 
${\qquad \;\;\;\gra (6)}\quad$ $\dfrac{1}{\sqrt{2}}\begin{pmatrix} 
-i & -1\\
-i & 1
\end{pmatrix}$ \\
${\gra (2)}$ $\dfrac{1}{\sqrt{2}}\begin{pmatrix} 
1 & 1\\
-i & i
\end{pmatrix}$  & 
${\gra (2)}\quad$ $\dfrac{1}{\sqrt{2}}\begin{pmatrix} 
1 & -i\\
1 & i
\end{pmatrix}$ &
 $\,{\gra (2)}\quad\;$ $\begin{pmatrix} 
0 & 1\\
1 & 0
\end{pmatrix}$ 
${\qquad \;\;\;\gra (7)}\quad$  $\dfrac{1}{\sqrt{2}}\begin{pmatrix} 
-i & 1\\
i & 1
\end{pmatrix}$\\
${\gra (3)}$ $\dfrac{1}{\sqrt{2}}\begin{pmatrix} 
1 & -i\\
i & -1
\end{pmatrix}$ &${\gra (3)}\quad$  $\dfrac{1}{\sqrt{2}}\begin{pmatrix}
1 & -i\\
i & -1
\end{pmatrix}$ &${\gra (3)}\quad$  $\begin{pmatrix} 
0 & -1\\
1 & 0
\end{pmatrix}$ 
${\qquad\;\; \gra (8)}\quad$  $\dfrac{1}{\sqrt{2}}\begin{pmatrix} 
i & i\\
-1 & 1
\end{pmatrix}$\\
${\gra (4)}\quad\quad$ $\begin{pmatrix} 
0 & -i\\
1 & 0
\end{pmatrix}$  &
 &${\,\gra (4)}\quad$  $\begin{pmatrix} 
1 & 0\\
0 & -1
\end{pmatrix}$
${\qquad\;\, \gra (9)}\quad$  $\dfrac{1}{\sqrt{2}}\begin{pmatrix} 
-i & -i\\
-1 & 1
\end{pmatrix}$\\
& &${\,\gra (5)}$  $\dfrac{1}{\sqrt{2}}\begin{pmatrix} 
i & -1\\
i & 1
\end{pmatrix}$
 $\qquad {\gra (10)}\;\;$  $\dfrac{1}{\sqrt{2}}\begin{pmatrix} 
-i & i\\
1 & 1
\end{pmatrix}$\\
\bottomrule
\end{tabular}
\end{table}
\end{center}

We experimentally implement both Alice's measurement operators and repreparation operators through a sequence of two WPs (QWP, then HWP), and Alice's projective measurement in a PBS measuring in $\left\lbrace \ket{H},\ket{V}\right\rbrace$.
Bob's operators are implemented  via three WPs (QWP, then HWP, then QWP). 
In Table \ref{tab:angle} we show the specific WP angles we use for each operator.

\begin{center}
\begin{table}[htb]
\caption{\footnotesize \textbf{Set of WPs angles.} A list of all of the WP angles used to perform the operators listed in Table \ref{tab:op}. In our experiment, all permutations of these settings were used, which, together with our three input states, results in $360$ measurement settings.\vspace{3mm}}
\label{tab:angle}
\centering
\begin{tabular}{l|l|l}
\toprule
{$\qquad$ \textbf{Alice - Meas.}} & $\quad$ {\textbf{Alice - Reprep.}} & $\qquad\quad$ {\textbf{Bob - Unit. Op.}} \\
\midrule
& &
$\,{\gra (1)}\quad$ $\text{0\degree}_{\text{QWP}} \,$, $\text{0\degree}_{\text{HWP}} \,$, $\text{0\degree}_{\text{QWP}}$ \\
${\gra (1)}\quad$ $\text{0\degree}_{\text{HWP}} \,$, $\text{0\degree}_{\text{QWP}}$ & 
${\gra (1)}\quad$ $\text{0\degree}_{\text{HWP}} \,$, $\text{0\degree}_{\text{QWP}}$ &
$\,{\gra (2)}\quad$ $\text{0\degree}_{\text{QWP}} \,$, $\text{45\degree}_{\text{HWP}} \,$, $\text{0\degree}_{\text{QWP}}$ \\
& &
$\,{\gra (3)}\quad$ $\text{90\degree}_{\text{QWP}} \,$, $\text{45\degree}_{\text{HWP}} \,$, $\text{0\degree}_{\text{QWP}}$\\
${\gra (2)}\quad$ $\text{22.5\degree}_{\text{HWP}} \,$, $\text{45\degree}_{\text{QWP}}$ & ${\gra (2)}\quad$ $\text{22.5\degree}_{\text{HWP}} \,$, $\text{0\degree}_{\text{QWP}}$ &
$\,{\gra (4)}\quad$ $\text{90\degree}_{\text{QWP}} \,$, $\text{0\degree}_{\text{HWP}} \,$, $\text{0\degree}_{\text{QWP}}$ \\
& & $\,{\gra (5)}\quad$ $\text{90\degree}_{\text{QWP}} \,$, $\text{0\degree}_{\text{HWP}} \,$, $\text{45\degree}_{\text{QWP}}$ \\
${\gra (3)}\quad$ $\text{0\degree}_{\text{HWP}} \,$, $\text{-45\degree}_{\text{QWP}}$ & 
${\gra (3)}\quad$ $\text{0\degree}_{\text{HWP}} \,$, $\text{-45\degree}_{\text{QWP}}$  & $\,{\gra (6)}\quad$ $\text{90\degree}_{\text{QWP}} \,$, $\text{45\degree}_{\text{HWP}} \,$, $\text{45\degree}_{\text{QWP}}$ \\
& & $\,{\gra (7)}\quad$ $\text{0\degree}_{\text{QWP}} \,$, $\text{0\degree}_{\text{HWP}} \,$, $\text{45\degree}_{\text{QWP}}$ \\
${\gra (4)}\quad$ $\text{45\degree}_{\text{HWP}} \,$, $\text{0\degree}_{\text{QWP}}$ & 
 & $\,{\gra (8)}\quad$ $\text{45\degree}_{\text{QWP}} \,$, $\text{0\degree}_{\text{HWP}} \,$, $\text{90\degree}_{\text{QWP}}$ \\
& & $\,{\gra (9)}\quad$ $\text{45\degree}_{\text{QWP}} \,$, $\text{45\degree}_{\text{HWP}} \,$, $\text{90\degree}_{\text{QWP}}$ \\
& & ${\gra (10)}\;\;$ $\text{45\degree}_{\text{QWP}} \,$, $\text{0\degree}_{\text{HWP}} \,$, $\text{0\degree}_{\text{QWP}}$\\
\bottomrule
\end{tabular}
\end{table}
\end{center}

\paragraph{Appendix III. Experimentally Estimating Probabilities}\vspace{3mm} 

Since Alice makes a two-outcome measurement and our final measurement has two outcomes, for each setting of Alice and Bob there are four different outcomes.
Experimentally, each outcome corresponds to a different single-photon detector. 
For each setting we collect approximately 7500 counts in total after 2 seconds of data acquisition.
From these counts we estimate the four corresponding output probabilities through the formula
\begin{equation}
\label{eq:counts}
p_{m n} = \frac{C_{m n}}{C_{\text{tot}}\cdot \eta_m \cdot \eta_{n}^m},
\end{equation}
where $C_{m n}$ is the number of counts collected at one of the detectors and the $\eta$ factors are different relative detector efficiencies, described below.
Here, $m$ labels Alice's outcome (experimentally, this labels in the internal (blue) or external (yellow) interferometer), and $n$ the outcome of the final measurement (experimentally, port 0 or port 1 of either interferometer).
The total number of (efficiency-corrected) counts, appearing in Eq. \ref{eq:counts}, is
\begin{equation}
C_{\text{tot}} = \sum_{m=0}^1 \sum_{n=0}^1 \dfrac{C_{m n}}{\eta_{m} \cdot \eta_{n}^m}.
\end{equation}

The efficiency factors in the above equations are defined as follows. 
The single-subscript factor $\eta_m$ refers to relative efficiencies between the internal ($m=1$) and external ($m=0$) interferometer (Figure \ref{img:experimental_apparatus1}).
The other factors $\eta_n^m$ refer to the relative efficiencies between the two ports $n=0$ and $n=1$, of interferometer $m$.
Then the absolute efficiency of a given detector is $\eta_m\cdot\eta_n^m$.
Roughly speaking, to estimate the relative efficiencies we must send the same number of photons between the detectors and compare the measured count rates.

\begin{figure}[hbt]
\centering
\includegraphics[width=.87\textwidth]{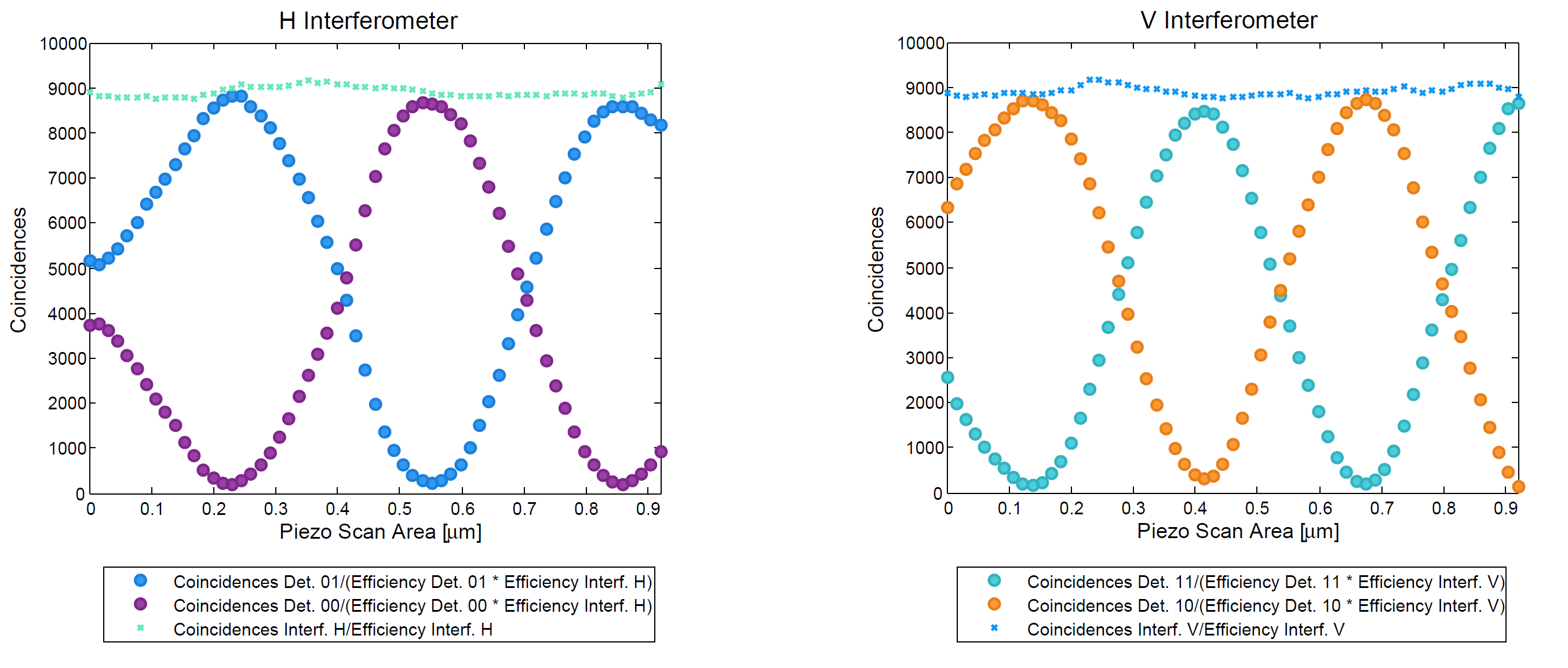}
\caption{\footnotesize \textbf{Efficiency-corrected interferometer fringes out of the two interferometers.} A plot of the coincidences between the herald and the two detectors at the output of each interferometers as the interferometer phase is varied.}
\label{img:visi}
\end{figure}

\begin{figure}[hbt]
\centering
\includegraphics[width=.87\textwidth]{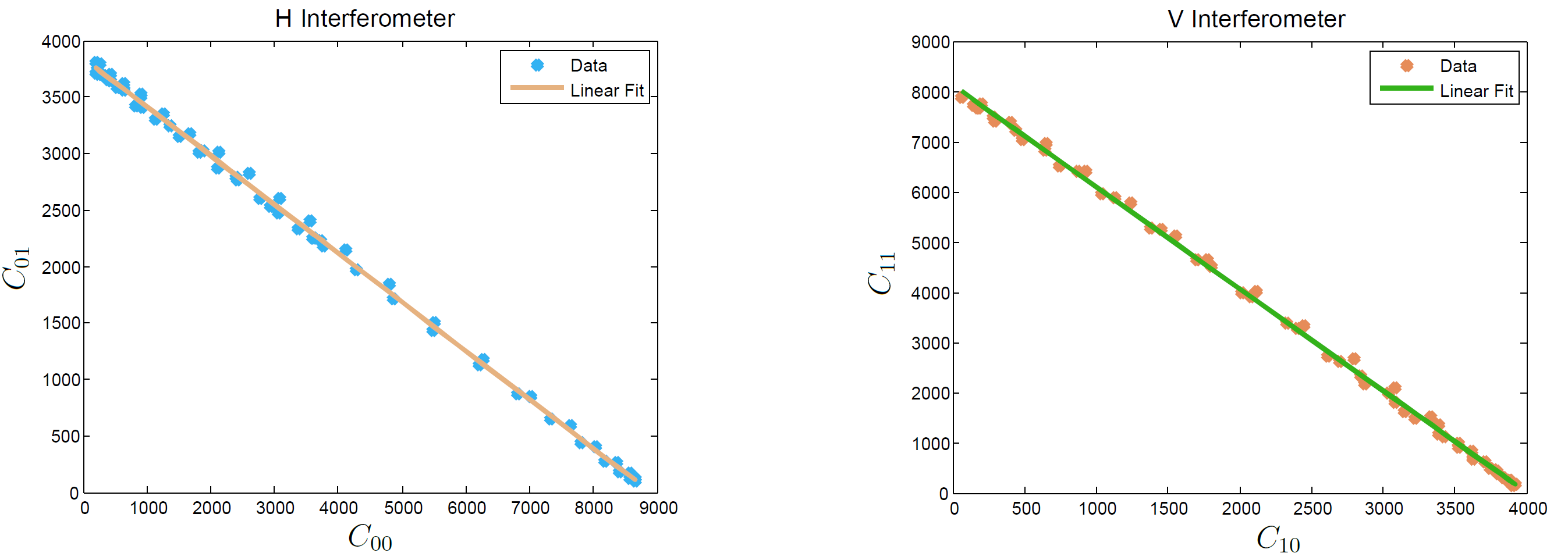}
\caption{\footnotesize \textbf{Determination of detection efficiency.} Triggered coincidences detected in port $1$ plotted against those detected in port $0$ for both interferometers. Since total number of photons exiting the interferometer should be constant, the relative collection/detection efficiency can be determined from the slope of this line.}
\label{img:Deph}
\end{figure}

To estimate $\eta_n^m$ within each interferometer we send the photons between the two ports by scanning the phase (when all of the internal WPs are set to $0$) by means of a piezo-electrically driven translation stage.
Plots of representative interference fringes (already efficiency corrected) for each interferometer are shown in Figure \ref{img:visi}.
By requiring the total counts out of each port to be constant, we can obtain a relative efficiency between the two ports in each interferometer.
In practice, we obtain the efficiency by plotting the counts out of one port versus the counts out of the other port.
If the two efficiencies are equal, the slope of this line will be $1$.
However, due to different coupling and detector efficiencies must enforce this by we requiring
\begin{equation}
\label{eq:eff}
K_0 = \dfrac{C_{00}}{\eta_{0}^0} + \dfrac{C_{01}}{\eta_{1}^0} \qquad \mathrm{and} \qquad K_1 = \dfrac{C_{10}}{\eta_{0}^1} + \dfrac{C_{11}}{\eta_{1}^1},
\end{equation}
where $K_0$ and $K_1$ are constants. 
We will set one efficiency of each pair to 1, since we are interested in the relative efficiency.
Setting (arbitrarily) $\eta_{0}^m=1$ means that the slope of $C_{m1}$ versus $C_{m0}$ will be $\eta_{1}^m$.
These plots, for both interferometers, are shown in Figure \ref{img:Deph}.

If we next estimate $\eta_m$, the relative efficiency between two interferometers, we are able to estimate the required probabilities (Eq. \ref{eq:counts}).
To do this we use the state preparation waveplate (Figure \ref{img:experimental_apparatus1}) to send the photons all to one interferometer or the other.
In each case we scan the phase.
Then, using the previously-discussed efficiencies $\eta_{n}^m$, we have $K_0$ and $K_1$ (Eq. \ref{eq:eff}).
As before, we can set one of the relative efficiencies to 1, we choose $\eta_0=1$.
Then we can calculate the final efficiency as
\begin{equation}
\eta_1 = \dfrac{\text{mean value}(K_1)}{\text{mean value}(K_0)}.
\end{equation}
This works because using the WPs and the PBS we can send nearly all of the incident photons one way or the other.

Using this procedure we now have relative efficiencies between all of the detectors.  
Note that $\eta_0\cdot\eta_0^0=1$; however, this does not matter as even if we had the absolute efficiency of each detector it would cancel out in the calculation of the probability (Eq. \ref{eq:counts}), because we must normalise by $C_\mathrm{tot}$.
After evaluating $p_{00}$, $p_{01}$, $p_{10}$, and $p_{11}$ for each of Alice and Bob's settings we weight each by the corresponding $\alpha_{a,d,x,y,z}$ (Eq. \ref{eqn:CNS}), and sum them all up.
This gives us our experimental value of the causal non-separability.

\paragraph{Appendix IV. Stability and Visibility of the Interferometers}\vspace{3mm} 
Central to our experiment are two interferometers whose overall size is approximately $80$ cm $\times$ $120$ cm. 
The visibility of the two interferometers is 95\%, this is apparent in the interferograms shown in Figure \ref{img:Deph}. 
This error can be interpreted as dephasing noise on the control qubit (see the discussion in the main text).

In addition to the reduced visibility, the phase of the interferometer fluctuates.
If the phase fluctuates on the the time scale of the acquisition time this would further decrease the visibility.
However, we find the phase drifts rather slowly, by approximately 0.01 rad in one minute.

To measure the causal witness we need to set $259$ different waveplate settings.  
Moving the waveplates from one setting to the next takes approximately 30 seconds. 
Combined with the measurement time of two seconds this means it takes approximately 30 seconds per measurement setting.
Therefore, after 30 measurements the phase drifts enough to cause a noticeable error.
To combat this, we automatically reset the phase to $0$ rad every 20 measurement settings by setting the WPs to $0^\circ$, scanning the piezo-electrically driven translation stage, and moving to the maximum of the fringe.

In spite of this there is still residual phase drift.  
To characterize the remaining error we simply repeated our experimental procedure (setting the waveplates, counting for 2 seconds, and resetting the phase every 20 measurements); however, we set the WPs to $0^\circ$ each time.
Thus we should have always been at a maximum or minimum of a fringe.
By measuring the deviation from the ideal values we estimate that, over the course of our entire data run, we have a residual phase fluctuation of approximately $0.04$ rad.
We then propagate this error in order to estimate an error on each probability that we measure.
These are the errorbars drawn in Figure \ref{img:first_70} and Supplementary Figures 1-3.

\paragraph{Appendix V. Causal Witness Derivation for our Set-Up}\vspace{3mm} 

Here we define what a `causal witness' is, and sketch the algorithm that was used to compute the witness suitable for our experimental set-up. We refer the reader to Ref.~\cite{NewJournPhys.17.102001} for an exhaustive introduction to the subject. Throughout this section we will use the Choi-Jamio\l{}kowski isomorphism, which we introduce briefly in the \textit{\textbf{Appendix}}, Sec. VI.

A process matrix (PM) $W^\text{sep} \in \mathcal{H}^{\text{(in)}} \otimes \mathcal{H}^A \otimes \mathcal{H}^B \otimes \mathcal{H}^{\text{(out)}}$ is `causally separable' if it can be written as convex combination of processes compatible with the causal order $A \rightarrow B$ and $B \rightarrow A$, that is, as ${W}^{\text{sep}} = p {W}^{A\rightarrow B} + (1-p){W}^{B\rightarrow A}$.

A `causal witness' $S \in \mathcal{H}^{\text{(in)}} \otimes \mathcal{H}^A \otimes \mathcal{H}^B \otimes \mathcal{H}^{\text{(out)}}$ is a Hermitian operator such that for all `causally non-separable' PMs
${W}^{n-sep}$, $\tr(S\, {W}^{n-sep}) < 0$, but for any process ${W}^{sep}$, $\tr(S\, {W}^{sep}) \geq 0$. The \textit{optimal} causal witness $S_\text{opt}$ for a given process $W$ can be computed efficiently using a `semidefinite program' (SDP) \cite{NewJournPhys.17.102001}:
\begin{equation}
\begin{gathered}
\min \tr(S\, W) \\
\text{s.t.}\quad S \in \mathcal S \qquad \mathcal{I}^{\mathcal{H}^\text{(in)} \otimes \mathcal{H}^A \otimes \mathcal{H}^B \otimes \mathcal{H}^\text{(out)}}/d_\text{out} - S \in \mathcal W^\ast
\end{gathered}
\end{equation}
where $\mathcal S$ and $\mathcal W^\ast$ are, respectively, the set of causal witnesses and set of Hermitian operators that have non-negative trace with process matrices, as defined in \cite{NewJournPhys.17.102001}, and $\mathcal{I}^{\mathcal{H}^\text{(in)} \otimes \mathcal{H}^A \otimes \mathcal{H}^B \otimes \mathcal{H}^\text{(out)}}/d_\text{out}$ is the identity operator on ${\mathcal{H}^\text{(in)} \otimes \mathcal{H}^A \otimes \mathcal{H}^B \otimes \mathcal{H}^\text{(out)}}$ divided by the dimension of the output spaces $d_{\text{out}} := d_{H^{\text{(in)}}} d_{\mathcal{H}^A_{\text{out}}} d_{\mathcal{H}^B_{\text{out}}}$ for normalization.

The causal non-separability $\text{CNS} ({W}^{n-sep}) = -\tr(S_\text{opt} {W}^{n-sep})$ is the minimal $\lambda \ge 0$ such that the PM 
\begin{equation}
W_\lambda = \frac1{1+\lambda}\big({W}^{n-sep}+\lambda\Omega\big)
\end{equation}
is causally separable, optimized over all valid process matrices $\Omega$. This means that it is the minimum amount of worst-case noise necessary to make ${W}^{n-sep}$ causally separable or, equivalently, the maximum (or rather the supremum) amount of worst-case noise that ${W}^{n-sep}$ can tolerate before becoming causally separable. Noting that $\frac{1}{1+\lambda} + \frac{\lambda}{1+\lambda} = 1$, we see that $\frac{\lambda}{1+\lambda}$ can be interpreted as the probability that the worst-case process $\Omega$ is prepared instead of the desired process $W^{n-sep}$, and therefore that $\frac{\text{CNS}(W)}{1+\text{CNS}(W)}$ is the maximal such probability that still allows us to see causal non-separability.

Any witness $S$ (in particular, $S_\text{opt}$) can be decomposed with respect to a \textit{basis} for the space $\mathcal{H}^{\text{(in)}} \otimes \mathcal{H}^A \otimes \mathcal{H}^B \otimes \mathcal{H}^{\text{(out)}}$. Such a basis consists of the CJ representations of general state preparation on $\mathcal{H}^{\text{(in)}}$, general measurement and repreparation operations on $\mathcal{H}^{A}$ and $\mathcal{H}^{B}$, and a general measurement on $\mathcal{H}^{\text{(out)}}$. Having access to such a basis of operations means being able to perform full `causal tomography'.

However, in our experimental set-up, Alice can implement general MR operations $M_{a, x}^A$, but Bob can implement only unitary operations $U_y^B$, and measurements are done only in the superposition basis. Thus $S_\text{opt}$ will not necessarily be experimentally achievable, and in our case it is not. To compute the best witness that we can experimentally implement, we add a restriction on the decomposition of the witness as an additional constraint in the SDP, which then outputs the optimal \textit{experimentally accessible} witness $S_{\text{exp}}$:
\begin{equation}\label{eqn:algorithm}
\begin{gathered}
\min \tr(S\, W) \\
\text{s.t.}\quad S \in \mathcal S \qquad \mathcal{I}^{\mathcal{H}^\text{(in)} \otimes \mathcal{H}^A \otimes \mathcal{H}^B \otimes \mathcal{H}^\text{(out)}}/d_\text{out} - S \in \mathcal W^\ast \\
S = \sum_{z=0}^2 \sum_{a=0}^{1} \sum_{x=0}^{11} \sum_{y=0}^{9} \sum_{d=0}^1 \alpha_{a,d,x,y,z} \cdot \rho^{(\text{in})}_z \otimes M_{a, x}^A \otimes U_y^B \otimes D^{(\text{out})}_d
\end{gathered}
\end{equation}
where $\bigl\lbrace M_{a, x}^A \bigr\rbrace$ are the $24$ CJ representations of measurement-repreparation maps, among which $16$ are linearly independent, $\bigl\lbrace U_y^B \bigr\rbrace$ the $10$ linearly independent CJ representations unitaries, which are listed in Sec. II, and $\{D^{(\text{out})}_d\}$ the two projectors onto the superposition basis.

The algorithm \ref{eqn:algorithm} returns the coefficients $\alpha_{a,d,x,y,z}$, which are used to weigh the experimental probabilities $p\bigl(a, d | x, y, z\bigr)$ corresponding to $\tr\bigl[\bigl(\rho^{(\text{in})}_z \otimes M_{a, x}^A \otimes U_{y}^B \otimes D^{(\text{out})}_d\bigr) \cdot W_{\text{SWITCH}} \bigr]$ to compute the experimental value for $\tr(S_{\text{exp}} W_{\text{SWITCH}})$.

Analogously to the ideal case, the `experimentally accessible causal non-separability' (\textit{i.e.}, $\text{CNS}_\text{exp} (W_{\text{SWITCH}}) = -\tr(S_\text{exp} W_{\text{SWITCH}})$) is the maximal amount of worst-case noise that can be admixed to $W_{\text{SWITCH}}$ before our experimental set-up becomes incapable of certifying that $W_{\text{SWITCH}}$ is causally non-separable, and $\frac{\text{CNS}_\text{exp}(W_{\text{SWITCH}})}{1+\text{CNS}_\text{exp}(W_{\text{SWITCH}})}$ the maximal probability of preparing the worst-case noise process instead of the ideal $W_{\text{SWITCH}}$.

\paragraph{Appendix VI. Choi-Jamio\l{}kowski Isomorphism}\vspace{3mm}
In this section we define the \textit{Choi-Jamio\l{}kowski isomorphism} \cite{CHOI1975285}, as used in the main text. For the sake of clarity, it is convenient to first introduce the isomorphism for unitary operators and then to generalize it to linear maps. 

\paragraph{\textit{Unitary operators.}} Consider an operator $U \in \mathcal{L}(\mathcal{H}_{pur}^{\text{in}})$ (where $\mathcal{L}(\mathcal{H}_{pur})$ is the space of linear operators in the Hilbert space of pure states $\mathcal{H}_{pur}$) such that $U : \mathcal{H}_{pur}^{\text{in}} \rightarrow \mathcal{H}_{pur}^{\text{out}}$. It is always possible to rewrite the operator $U$ acting on the basis $\lbrace\ket{j}\rbrace \in \mathcal{H}_{pur}^{\text{in}}$ into a new basis $\lbrace\ket{k}\rbrace \in\mathcal{H}_{pur}^{\text{out}}$
\begin{equation}
U  = \sum_{j,k} U_{k,j} \ket{k}\bra{j}
\label{eqn:ini}
\end{equation}
where $U_{k,j} = \bra{k}U\ket{j}$.

According to the Choi-Jamio\l{}kowski isomorphism (CJ), a linear operator acting on a Hilbert space $\mathcal{H}_{pur}^{\text{in}}$ is isomorphic to a vector in $\mathcal{H}_{pur}^{\text{in}} \otimes \mathcal{H}_{pur}^{\text{out}}$ \cite{PhysRevA.88.052130, NewJournPhys.17.102001}:
\begin{equation}
\text{CJ} : \mathcal{L}(\mathcal{H}_{pur}^{\text{in}}) \rightarrow \mathcal{H}_{pur}^{\text{in}}\otimes\mathcal{H}_{pur}^{\text{out}}
\end{equation}

We define a (non-normalized) maximally entangled state via the double-ket notation, as 
\begin{equation}
\ket{\mathbb{1}\rangle}^{\text{in, in}} = \sum_{j} \ket{j}\otimes\ket{j} \in \mathcal{H}_{pur}^{\text{in}} \otimes \mathcal{H}_{pur}^{\text{in}}
\end{equation}
The application of the CJ isomorphism on the linear operator $U$ gives 
\begin{equation}
\text{CJ}(U) = \bigl(\mathbb{1}\otimes U^{\ast}\bigr)\ket{\mathbb{1}\rangle}^{\text{in, in}} = \sum_{j} \ket{j}\otimes U^{\ast}\ket{j} = \sum_{j,k} U_{k,j}^{\ast} \ket{j}\otimes \ket{k} := \ket{U^{\ast}\rangle}^{\text{in, out}}
\label{eqn:fin}
\end{equation}
where $U^{\ast}$ is the complex conjugate and $\ket{U^{\ast}\rangle}$ is the CJ representation (or CJ vector) of $U$.

By comparing Eq. \ref{eqn:ini} and \ref{eqn:fin}, one can see the direct correspondence between the operator $\ket{k}\bra{j}$ and the vector $\ket{j}\otimes \ket{k}$.  Thus, a unitary operator is represented as a pure state in the CJ isomorphism.

\paragraph{\textit{Linear maps.}} Consider a linear map $\mathcal{M} : \mathcal{H}^{\text{in}} \rightarrow \mathcal{H}^{\text{out}}$ (where $\mathcal{H}$ is the Hilbert space of the density matrices, \textit{i.e.}, $\mathcal{L}(\mathcal{H}_{pur})$). The corresponding CJ representation is
\begin{equation}
M^{\text{in, out}} := \bigl[\mathcal{I} \otimes \mathcal{M} \bigl(\ket{\mathbb{1}\rangle}\bra{\langle \mathbb{1}}\bigr)\bigr]^T
\end{equation} 
This equation can be led back to Eq. \ref{eqn:fin} when $\mathcal{M} (\rho) = U \rho U^{\dagger}$:
\begin{align}
M^{\text{in, out}} :&= \left[\mathcal{I} \otimes U \left(\sum_{j, j_1} \ket{j}\otimes \ket{j} \bra{j_1}\otimes \bra{j_1}\right) U^{\dagger}\right]^T =\notag\\
&= \left[\sum_{j, j_1, k, k_1} U_{k,j}^{\ast}U_{j_1,k_1} \ket{j} \otimes \ket{k} \bra{j_1} \otimes \bra{k_1}\right]^T = \ket{{U^{\ast}}\rangle} \bra{\langle{U^{\ast}}}
\end{align} 
Thus, an arbitrary linear map is represented by a matrix in the CJ isomorphism.

By looking at the action of a linear operator, such as a time evolution $\ket{j} \mapsto \ket{k}$, it turns out that in the CJ notation it is not required to  ascribe a temporal order to the application of different operators.
Consequently, analysing situations in which the causal order of events is not well-defined (such as in the quantum SWITCH) in the CJ isomorphism is very convenient.

\newpage

\begin{suppfigure}[hbt]
\centering
\includegraphics[width=.87\textwidth]{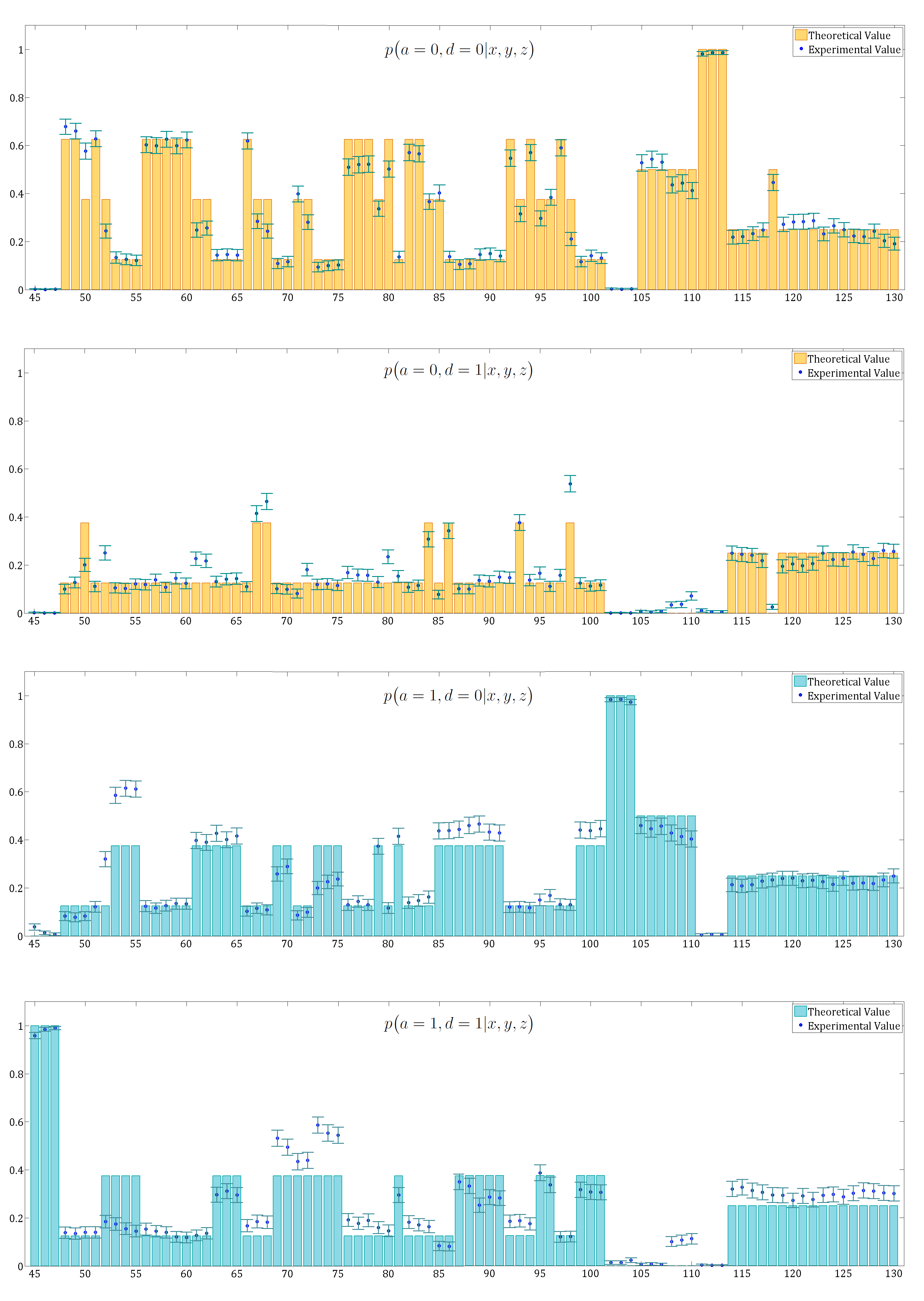}
\caption{\footnotesize \textbf{Experimentally estimated probabilities.} Continuation of Figure \ref{img:first_70}.}
\label{img:det-23}
\end{suppfigure}

\begin{suppfigure}[hbt]
\centering
\includegraphics[width=.86\textwidth]{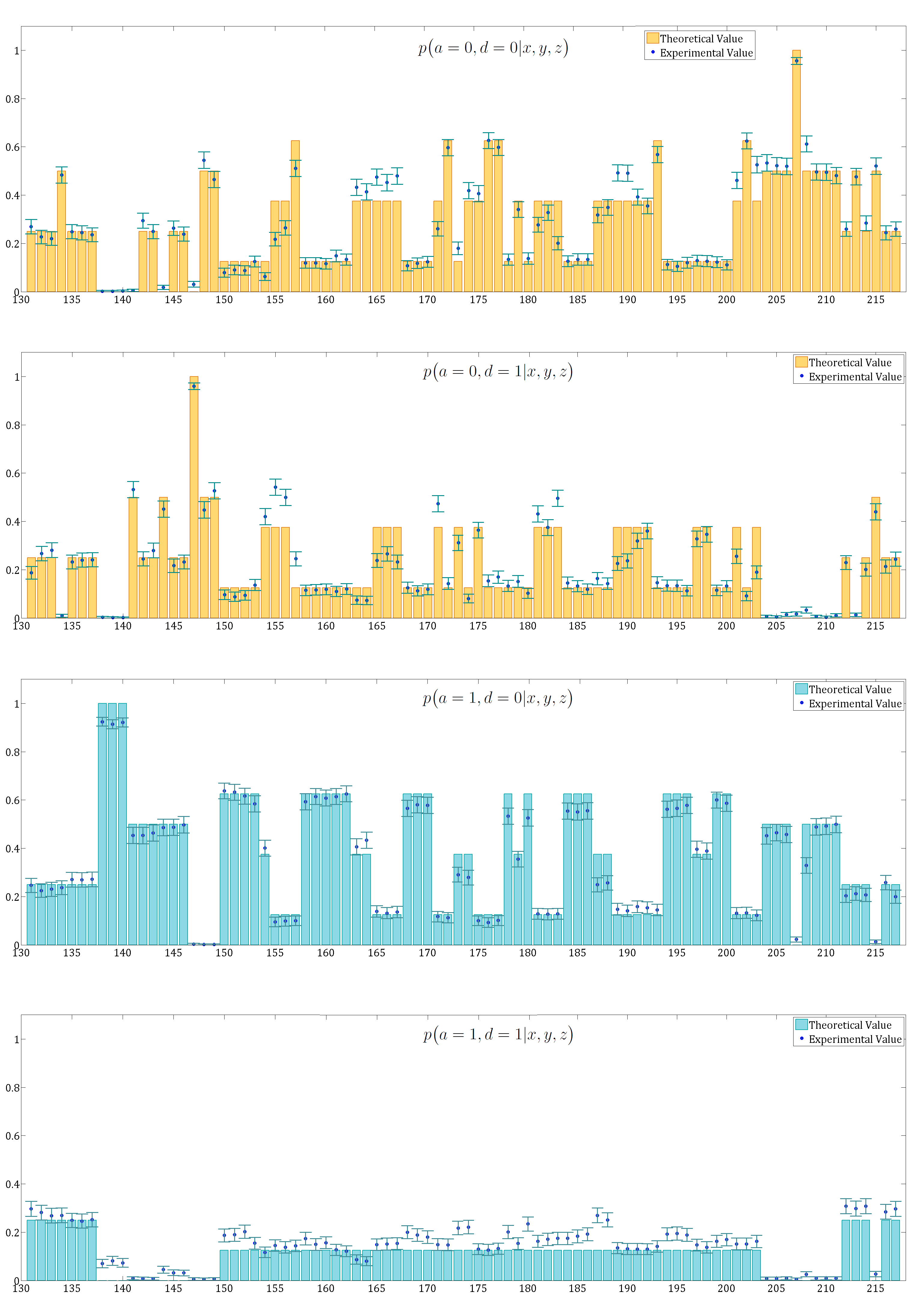}
\caption{\footnotesize \textbf{Experimentally estimated probabilities.} Continuation of Figure \ref{img:first_70}.}
\label{img:det-45}
\end{suppfigure}

\begin{suppfigure}[hbt]
\centering
\includegraphics[width=.87\textwidth]{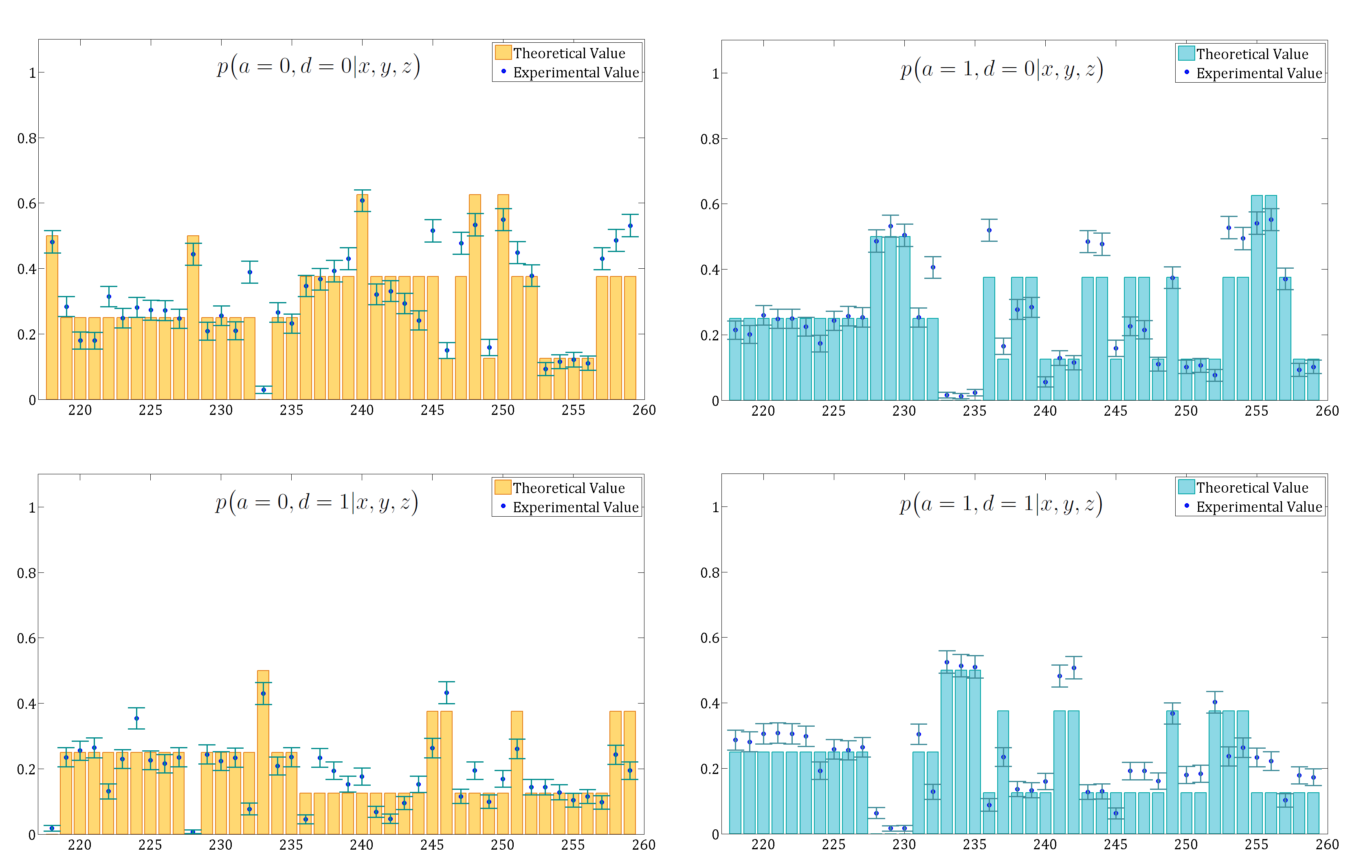}
\caption{\footnotesize \textbf{Experimentally estimated probabilities.} Continuation of Figure \ref{img:first_70}.}
\label{img:det-6}
\end{suppfigure}


\end{document}